\documentclass[letterpaper,times]{IONconf}

\usepackage{times}
\usepackage{epsfig}
\usepackage[T1]{fontenc}
\usepackage{graphicx}
\usepackage{booktabs}
\usepackage[misc]{ifsym}

\usepackage{paralist}
\usepackage{eurosym}
\usepackage{flushend}
\usepackage{amsmath}
\usepackage{amssymb}
\usepackage{url}
\usepackage[utf8]{inputenc}
\usepackage{amsfonts}
\usepackage{soul}
\usepackage[justification=centering]{caption}
\usepackage[list=true]{subcaption}
\usepackage{tabu}
\usepackage{wrapfig}
\usepackage{arydshln}
\usepackage[normalem]{ulem}
\usepackage{multirow}
\usepackage{siunitx}
\usepackage{natbib}
\usepackage{hyperref}
\usepackage{makecell}
\usepackage{adjustbox}
\usepackage{array}
\usepackage{diagbox}
\usepackage{tikz}
\usetikzlibrary{arrows.meta,positioning,calc,fit,backgrounds,shapes.geometric,shadows}

\title{Analyzing and Characterizing Multi-Source Interference Effects at Jammertest Norway 2025}

\author{Lucas Heublein,\,
{I\~{n}igo} Cort\'{e}s Vidal,\,
Tobias Feigl,\,
Alexander Rügamer,\,
Felix Ott\\
\textit{Fraunhofer Institute for Integrated Circuits IIS, Nürnberg, Germany}\\
{\tt\small\{lucas.heublein, tobias.feigl, alexander.ruegamer, felix.ott@\}iis.fraunhofer.de}}

\begin{document}

\maketitle

\section*{BIOGRAPHY}

\biography{Lucas Heublein} earned his M.Sc.~in Integrated Life Science from Friedrich-Alexander University (FAU) Erlangen-Nürnberg and completed his Master of Science in Computer Science in 2024. He began his association with the Hybrid Positioning \& Information Fusion group at the Fraunhofer Institute for Integrated Circuits IIS in 2020 as a student assistant and currently serves as a research assistant in the Self-Learning Systems group.

\biography{I\~{n}igo Cort\'{e}s Vidal} received his M.Sc.~degree in telecommunication engineering from the School of Engineering at San Sebasti\'{a}n, University of Navarra, Spain, in April 2018. Since then, he has been working in the Satellite-based Localization Systems department at Fraunhofer IIS, Nürnberg, Germany. In September 2022, he received his Ph.D.~in computing and electrical engineering from Tampere University, Finland.

\biography{Tobias Feigl} received his Ph.D.~degree in Computer Science from the Friedrich-Alexander University (FAU) Erlangen-Nürnberg in 2021 and his Masters degree from the University of Applied Sciences Erlangen-Nuremberg, Germany, in 2017. In 2017 he joined the Fraunhofer IIS Nuremberg, Germany, where he worked as a student since 2009, with a strong focus on AI and information fusion. In parallel, since 2017 he is a lecturer at the Computer Science department at FAU, where he gives courses on machine and deep learning and supervises related qualification work.

\biography{Alexander Rügamer} received his Dr.-Ing.~degree from the FAU Erlangen-Nürnberg in 2024 and his Dipl.-Ing.~(FH) degree in Electrical Engineering from the University of Applied Sciences Würzburg-Schweinfurt in 2007. Since then, he has been working at the Fraunhofer IIS in the field of GNSS receiver development. From 2013 to 2023, he led a research group on secure GNSS receivers. Since 2024, he has been the head of the department of satellite-based localization. His main research interests focus on GNSS multi-band reception, integrated circuits, and immunity to interference.

\biography{Felix Ott}{received his M.Sc.~degree in Computational Engineering from FAU Erlangen-Nürnberg in 2019. He subsequently joined the Self-Learning Systems group within the Machine Intelligence department at the Fraunhofer IIS. In 2023, he was awarded his Ph.D.~from Ludwig-Maximilians University (LMU) Munich. He is currently managing the program Signals Intelligence. His research focuses on representation learning, domain adaptation, and few-shot learning for GNSS-based signals interference monitoring, such as interference signal characterization and direction finding.}

\section*{ABSTRACT}

Intentional radio-frequency interference from low-cost GNSS jammers increasingly threatens the accuracy and reliability of satellite-based positioning. Mitigating this threat requires not only detection but also robust waveform classification and characterization, direction-of-arrival inference for localization, and impact estimation on receiver performance under realistic operating conditions; all of which are challenged by strong distribution shifts across devices, sensors, environments, and satellite geometries. We address these challenges by compiling a dedicated real-world dataset recorded during Jammertest 2025 (And{\o}ya, Norway), covering two outdoor test areas with parallel measurements from a single-antenna E1/E5 receiver module and a $2 \times 2$ CRPA-array. The dataset spans diverse jamming, spoofing, and meaconing scenarios, including CW, PRN, sweep/chirp, and multi-emitter configurations, and is complemented by per-recording metadata enabling time-aligned ground truth. Methodologically, we benchmark 17 machine learning (ML) architectures for interference-modulation recognition and multi-task characterization (type, occupied bandwidth, and signal strength), and we quantify navigation-relevant degradation via a receiver-aware spectral separation coefficient (SSC) that maps measured spectra to effective $(C_s/N_0)_{\mathrm{eff}}$ loss. To enable multi-source analysis, we transfer a YOLOv8s and RF-DETR detector pretrained on a labeled spectrogram dataset to localize multiple simultaneous interferers in GNSS spectrograms and subsequently characterize each detected component. Results show near-99\% within-area classification accuracy but substantial cross-area performance drops, highlighting the need for robustness to real-world domain shifts. Overall, the proposed dataset and end-to-end, impact-aware pipeline provide a practical foundation for scalable multi-source interference monitoring and robust GNSS navigation in the field.

\textbf{Dataset available at:} \href{https://gitlab.cc-asp.fraunhofer.de/darcy_gnss/fraunhoferIIS_jammertest2025}{https://gitlab.cc-asp.fraunhofer.de/darcy\_gnss/fraunhoferIIS\_jammertest2025} \\ \textbf{Alternative download link: }\href{https://zenodo.org/records/21332689}{https://zenodo.org/records/21332689}
\section{INTRODUCTION}
\label{label_introduction}

Interference from intentional jamming devices can severely degrade the accuracy and reliability of GNSS-based positioning, and the problem has intensified with the growing availability of low-cost, easily accessible jammers~\citep{li_huang_lang,ferre_fuente,mehr_dovis}. Effective interference management requires not only detecting interference but also classifying and characterizing its waveform~\citep{heublein_feigl_jispin,clements_kriezis} (often a distinctive \textit{fingerprint} of the transmitter) and determining the direction-of-arrival to enable localization~\citep{heublein_wielenberg} and mitigation of the interference source.

Beyond source identification, robust navigation requires quantifying how interference translates into receiver-level performance loss: different jammer waveforms with identical total power can induce markedly different degradations depending on their spectral structure and overlap with the GNSS signal after front-end filtering. Therefore, interference monitoring must estimate not only the presence and class of a disturbance, but also its navigation-relevant impact in terms of correlator signal-to-noise-and-interference-ratio (SNIR) and the resulting reduction in effective carrier-to-noise density ratio $(C/N_0)_{\mathrm{eff}}$, which directly governs acquisition sensitivity, tracking stability, and demodulation robustness~\citep{betz_goldstein,bacic_sugar,enneking_antreich}. This motivates computing a physically grounded overlap metric such as the (receiver-aware) spectral separation coefficient (SSC), which links observable spectral characteristics and received interference power to expected $(C/N_0)_{\mathrm{eff}}$ degradation and thus enables comparable, receiver-specific impact assessment across heterogeneous interference types~\citep{kaplan}.

However, computing interference impact in practice faces several challenges: real-world data exhibit strong variability across jammer types, sensor characteristics, data formats (e.g., IQ samples, spectrograms, snapshot vs.~low-cost time-series data), jammer-receiver geometries and distances, changing satellite constellations, and environmental effects such as multipath and obstruction~\citep{heublein_feigl_crpa}. These distribution shifts and heterogeneous measurement conditions complicate the development of reliable, generalizable models and limit the transferability of methods validated only on synthetic or laboratory data~\citep{wielenberg_heublein_sionna}.

To address these challenges, we aim to train resilient ML methods for automated analysis, classification, and characterization of GNSS interference sources, including directional inference, and to support robust impact assessment under realistic operating conditions. We further extend the characterization to multi-interference scenarios~\citep{fonseca_santos,lin_zhang_tian,uvaydov_zhang} by detecting and separating multiple simultaneous interferers within a single spectrogram and estimating their individual attributes (type, occupied bandwidth, and signal strength) for per-source impact analysis. To enable this, we compile and provide a dedicated real-world dataset recorded during the Jammertest 2025 campaign in Norway (Andøya)~\citep{jammertest_gerard,johansson_spanghero}.

\textbf{Contributions.} Our main contributions are as follows. (1) A multi-day, multi-site real-world GNSS interference dataset from Jammertest 2025 with parallel measurements from heterogeneous receiver/antenna setups and time-resolved metadata for supervised benchmarking. (2) A systematic benchmark of 17 time-series ML architectures for interference-modulation recognition and multi-task characterization (type, bandwidth, signal strength), highlighting strong cross-area domain shifts. (3) A physically grounded, receiver-aware SSC-based impact evaluation that relates measured interference spectra and power to expected degradation in $(C_s/N_0)_{\mathrm{eff}}$. (4) A transfer-learning pipeline using a YOLOv8s~\citep{wang_yeh_liao} and RF-DETR~\citep{robinson_robicheaux} spectrogram detector to localize multiple simultaneous interferers and enable subsequent per-source characterization.
\section{RELATED WORK}
\label{label_related_work}

\paragraph{Interference Characterization} GNSS interference characterization has progressed from detection and class prediction toward richer, physically meaningful descriptions of interfering signals. Early ML-based monitoring methods, such as the twin SVM approach of \cite{li_huang_lang}, demonstrated real-time GNSS interference detection and classification, while \cite{ferre_fuente} and \cite{mehr_dovis} further showed the effectiveness of classical ML and CNN-based methods for jammer classification. More recent work has emphasized robustness and generalization under realistic data discrepancies: \cite{heublein_raichur_ion} studied supervised and unsupervised ML methods under real-world domain shifts, and \cite{heublein_feigl_crpa} extended this perspective to GNSS interference classification, characterization, and localization, motivating the joint estimation of jammer type, bandwidth, signal strength, and spatial information. Related studies also investigated attention-based fusion of IQ, FFT spectrogram, and AoA features for localization~\citep{heublein_wielenberg}, generative and variational models for compressed/disentangled GNSS spectrogram representations~\citep{heublein_feigl_jispin}, and simulation-based dynamic interference characterization and direction finding with the S-ICDF dataset~\citep{wielenberg_heublein_sionna}. In parallel, Jammertest has been introduced as an open GNSS interference test arena for resilient GNSS development~\citep{jammertest_gerard}, and recent real-world case studies underline the importance of detecting, characterizing, and identifying powerful interference sources~\citep{clements_kriezis,blanch_lo_chen,broumandan_pirsiavash}. Building on these works, our paper contributes a multi-site Jammertest 2025 dataset and an impact-aware characterization pipeline that jointly predicts waveform type, occupied bandwidth, and signal strength, separates simultaneous interferers in spectrograms, and links per-source estimates to receiver-aware SSC-based degradation metrics.

\paragraph{SSC for Impact Evaluation} The spectral impact of GNSS interference is commonly described by overlap-based receiver models. \cite{betz} showed that tracking degradation depends on interference power and spectral placement, while \cite{kaplan} formalized this through correlator-output SNIR and the SSC. \cite{betz_goldstein} related spectral overlap to jamming resistance for signal design. Later work emphasized receiver-specific interpretation: \cite{ruegamer_dissertation} considered front-end filtering and bandwidth effects, \cite{bacic_sugar} studied interference degradation in static GNSS measurements, and \cite{enneking_antreich} showed that SSC does not capture all tracking distortions. \cite{heublein_benschuh} verified receiver-aware SSC under interference-induced degradation. We build on these works by integrating receiver-aware SSC with ML-based interference characterization and multi-source spectrogram detection to map measured interference power and spectral overlap to effective $(C_s/N_0)_{\mathrm{eff}}$ loss in real Jammertest 2025 recordings.

\paragraph{Multi-Source Interference Detection} \cite{fonseca_santos} formulated radio-access-technology characterization as an object-detection problem on spectrograms, enabling time--frequency localization and bandwidth extraction even under overlapping transmissions. \cite{lin_zhang_tian} proposed a multi-signal detection framework for estimating carrier frequency and bandwidth from RF spectrum measurements. \cite{uvaydov_zhang} introduced semantic spectrum segmentation to classify and localize multiple signals with pixel-level granularity, arguing that segmentation can overcome limitations of rectangular bounding boxes. Our work uses multi-source detection for GNSS interference monitoring and per-source characterization.
\section{METHODOLOGY}
\label{label_methodology}

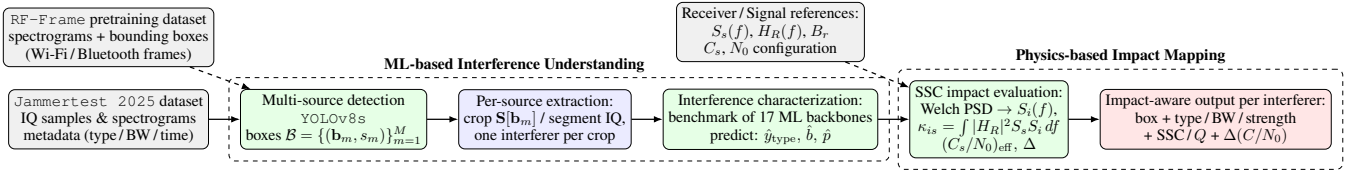
\begin{figure*}[!t]
\centering
\resizebox{\textwidth}{!}{%
\begin{tikzpicture}[
    node distance=0.75cm and 1.0cm,
    block/.style={
        rectangle,
        draw,
        rounded corners,
        align=center,
        minimum width=3.2cm,
        minimum height=0.9cm,
        fill=blue!8
    },
    data/.style={
        rectangle,
        draw,
        rounded corners,
        align=center,
        minimum width=3.2cm,
        minimum height=0.9cm,
        fill=gray!12
    },
    kernel/.style={
        rectangle,
        draw,
        rounded corners,
        align=center,
        minimum width=3.4cm,
        minimum height=0.95cm,
        fill=green!10
    },
    loss/.style={
        rectangle,
        draw,
        rounded corners,
        align=center,
        minimum width=3.4cm,
        minimum height=0.95cm,
        fill=red!10
    },
    arrow/.style={
        -{Latex[length=2mm]},
        thick
    }
]

\node[data] (jt) {\texttt{Jammertest 2025} dataset\\IQ samples \& spectrograms\\metadata (type\,/\,BW\,/\,time)};
\node[data, above=0.5cm of jt] (fordatis) {\texttt{RF-Frame} pretraining dataset\\spectrograms + bounding boxes\\(Wi-Fi\,/\,Bluetooth frames)};

\node[kernel, right=0.7cm of jt] (yolo) {Multi-source detection\\\texttt{YOLOv8s}\\boxes $\mathcal{B}=\{(\mathbf{b}_m,s_m)\}_{m=1}^{M}$};

\node[block, right=0.7cm of yolo] (crop) {Per-source extraction:\\crop $\mathbf{S}[\mathbf{b}_m]$ / segment IQ,\\one interferer per crop};

\node[kernel, right=0.7cm of crop] (char) {Interference characterization:\\benchmark of 17 ML backbones\\predict: $\hat{y}_{\mathrm{type}},\,\hat{b},\,\hat{p}$};

\node[data, above=0.6cm of char] (refs) {Receiver\,/\,Signal references:\\$S_s(f)$, $H_R(f)$, $B_r$\\$C_s$, $N_0$ configuration};

\node[kernel, right=0.7cm of char] (ssc) {SSC impact evaluation:\\Welch PSD $\rightarrow S_i(f)$,\\$\kappa_{is}=\int |H_R|^2 S_s S_i\,df$\\$(C_s/N_0)_{\mathrm{eff}},\,\Delta$};

\node[loss, right=0.7cm of ssc] (out) {Impact-aware output per interferer:\\box + type\,/\,BW\,/\,strength\\+ SSC\,/\,$Q$ + $\Delta(C/N_0)$};

\draw[arrow] (jt) -- (yolo);
\draw[arrow] (yolo) -- (crop);
\draw[arrow] (crop) -- (char);
\draw[arrow] (char) -- (ssc);
\draw[arrow] (ssc) -- (out);

\draw[arrow, dashed] (fordatis) -- (yolo);
\draw[arrow, dashed] (refs) -- (ssc);

\node[
    draw,
    dashed,
    rounded corners,
    fit=(yolo)(crop)(char),
    inner sep=0.25cm,
    label={[xshift=-1.0cm]above:\textbf{ML-based Interference Understanding}}
] {};

\node[
    draw,
    dashed,
    rounded corners,
    fit=(ssc)(out),
    inner sep=0.25cm,
    label=above:\textbf{{Physics-based Impact Mapping}}
] {};

\end{tikzpicture}
}
\caption{\textbf{Overview of the proposed methodology.} Jammertest recordings are processed to (i) detect multiple simultaneous interferers in spectrograms, (ii) characterize each detected interferer (type, occupied bandwidth, signal strength), and (iii) map spectral overlap and power to navigation-relevant degradation via a receiver-aware SSC pipeline.}
\label{figure_method_overview}
\end{figure*}

This section presents our methodology for impact-aware GNSS interference monitoring. Figure~\ref{figure_method_overview} visualizes our end-to-end pipeline. Section~\ref{label_method_characterization} introduces our interference characterization pipeline, where we benchmark time-series ML backbones and train a multi-task model to jointly predict interference type, occupied bandwidth, and signal strength. Section~\ref{label_method_ssc} derives and details the receiver-aware spectral separation coefficient (SSC) computation and its mapping to an effective carrier-to-noise density ratio $(C_s/N_0)_{\mathrm{eff}}$, enabling a physically interpretable estimate of navigation-relevant degradation. Section~\ref{label_method_multi_source} describes our multi-source interference detection approach, where a YOLOv8s-based detector localizes multiple simultaneous interferers in spectrograms and routes each detected component to the downstream characterization and impact assessment modules.

\subsection{Interference Characterization}
\label{label_method_characterization}

We benchmark a suite of 17 ML methods for interference-modulation recognition, spanning recurrent, convolutional, hybrid, and attention/MLP-based time-series architectures. Specifically, the evaluated backbones comprise LSTM and GRU (recurrent sequence models), FCN, ResNet, and ResCNN (1D convolutional baselines), the hybrid architectures LSTM-FCN, GRU-FCN, and MLSTM-FCN (recurrent--convolutional feature extractors), TCN (temporal convolutional network), multi-scale convolutional families InceptionTime and XceptionTime, OmniScale (multi-resolution temporal modeling), and more recent architectures including TST (time-series transformer), XCM, gMLP, and TSPerceiver. We train on the combined Area1+2 and select the best-performing method. Next, we train on labeled segments derived from the campaign metadata and evaluate generalization across test areas (Area~1, Area~2, and the combined Area~1+2 setting), thereby explicitly probing distribution shifts induced by changing environments, receiver conditions, and jammer configurations.

Our characterization task goes beyond waveform identification and is formulated as a multi-task learning problem in which each input segment is mapped to a structured output describing the interference instance. Concretely, given an input representation $\mathbf{x}$ (either raw IQ, magnitude/envelope, or a time--frequency feature such as a spectrogram slice), the model predicts (i) a discrete modulation/waveform class $\hat{y}_{\mathrm{type}}$, (ii) an occupied bandwidth estimate $\hat{b}$ (in MHz), and (iii) a received interference-strength estimate $\hat{p}$ (in dB). We implement this with a shared feature encoder $f_\theta(\cdot)$ and three task-specific heads,
\begin{equation}
    \mathbf{z}=f_\theta(\mathbf{x}), \qquad
    \hat{y}_{\mathrm{type}} = g_{\phi}(\mathbf{z}), \qquad
    \hat{b} = h_{\psi}(\mathbf{z}), \qquad
    \hat{p} = r_{\omega}(\mathbf{z}),
\end{equation}
where $g_\phi$ outputs class probabilities via a softmax layer and $(h_\psi,r_\omega)$ are regression heads. Training minimizes a weighted sum of losses,
\begin{equation}
\label{eq:multitask_loss}
    \mathcal{L}
    =
    \lambda_{\mathrm{cls}}\,\mathcal{L}_{\mathrm{CE}}\!\left(y_{\mathrm{type}}, \hat{y}_{\mathrm{type}}\right)
    +
    \lambda_{b}\,\lVert b-\hat{b}\rVert_{2}
    +
    \lambda_{p}\,\lVert p-\hat{p}\rVert_{2},
\end{equation}
where $\mathcal{L}_{\mathrm{CE}}$ denotes cross-entropy for modulation classification and the regression terms use an $\ell_2$ MSE-error for robustness against outliers. This formulation enables the model to learn a representation that is simultaneously discriminative for waveform type and calibrated to physically meaningful attributes relevant for downstream impact evaluation (e.g., SSC computation), rather than optimizing classification accuracy alone.

To reduce label noise and focus learning on operationally relevant disturbances, we additionally evaluate a thresholding strategy in which very weak jammers below a predefined power level are excluded from training and/or evaluation. We compare this multi-task formulation against a single-task classifier trained only with $\mathcal{L}_{\mathrm{CE}}$, allowing us to assess whether jointly predicting bandwidth and signal strength improves robustness and cross-area transfer. Performance is reported using modulation-type accuracy for the classification head and absolute errors for bandwidth (MHz) and signal strength (dB), providing a unified view of how well each model supports interference awareness and subsequent impact assessment.

\subsection{Spectral Separation Coefficient (SSC) for Impact Evaluation}
\label{label_method_ssc}

To quantify how each recorded interference source degrades the usable satellite information, we estimate a receiver-aware SSC and map it to an effective carrier-to-noise density ratio $(C_s/N_0)_{\mathrm{eff}}$, which directly governs acquisition sensitivity, tracking robustness, and demodulation performance. Concretely, we operate on complex baseband in-phase and quadrature (IQ) samples and model each snapshot as
\begin{equation}
    x[n] = s[n] + i[n] + w[n], \qquad n=1,\dots,N,
\end{equation}
where $s[n]$ denotes the desired GNSS signal, $i[n]$ the aggregate interference, and $w[n]$ additive noise. Under standard correlator assumptions, the prompt correlator output SNIR reveals that interference enters through a spectrally weighted overlap term between the desired signal and interference spectra. This motivates characterizing impact via the (receiver-aware) SSC
\begin{equation}
\label{eq:ssc-receiver-aware-method}
    \kappa_{is}
    =
    \int_{-\infty}^{\infty} \left|H_R(f)\right|^2 S_s(f) S_i(f)\, df,
\end{equation}
where $H_R(f)$ denotes the receiver transfer function, $S_s(f)$ is the normalized PSD of the desired signal, and $S_i(f)$ is the normalized PSD of the interference. In practice, the integration is restricted to the effective receiver bandwidth $[-B_r/2,B_r/2]$ (here $B_r=40.5\,\mathrm{MHz}$ due to digital filtering, despite a wider analog front-end), consistent with the interpretation of the SSC as a receiver-dependent overlap measure~\citep{betz,kaplan}. 

\textbf{PSD Estimation and Normalization.} For each recording we estimate spectra using a Welch periodogram. Let $\widehat{P}_{\mathrm{jam}}(f)$ denote the PSD estimate from a jammed interval and $\widehat{P}_{\mathrm{clean}}(f)$ the PSD estimate from an interference-free interval. To reduce residual GNSS signal contributions in the jammer spectrum, we form a non-negative difference spectrum
\begin{equation}
\label{eq:diff_psd}
    \widehat{P}_i(f) = \max\big\{\widehat{P}_{\mathrm{jam}}(f)-\widehat{P}_{\mathrm{clean}}(f),\,0\big\},
\end{equation}
and normalize it over the effective bandwidth to obtain
\begin{equation}
\label{eq:Si_norm}
    S_i(f) = \frac{\widehat{P}_i(f)}{\int_{-B_r/2}^{B_r/2}\widehat{P}_i(\nu)\, d\nu},
    \qquad
    \int_{-B_r/2}^{B_r/2} S_i(f)\, df = 1.
\end{equation}
For $S_s(f)$ we use the standard normalized PSD of the corresponding GNSS signal (GPS L1 or Galileo E1 B/C), and, if available, incorporate receiver filtering via $|H_R(f)|^2$ (otherwise $H_R(f)$ is approximated as unity within $[-B_r/2,B_r/2]$ and zero outside). With a discrete frequency grid $\{f_k\}_{k=1}^{K}$ and spacing $\Delta f$, the SSC is computed numerically as
\begin{equation}
\label{eq:ssc_discrete}
    \widehat{\kappa}_{is}
    =
    \sum_{k=1}^{K}
    \left|H_R(f_k)\right|^2
    S_s(f_k)\, S_i(f_k)\, \Delta f.
\end{equation}

\textbf{From SSC to Navigation-Relevant Degradation.} According to \cite{betz_goldstein}, the SSC induces the jamming resistance quality factor $Q$; for normalized $S_s(f)$ this simplifies to
\begin{equation}
\label{eq:Q_est}
    \widehat{Q} \approx \frac{1}{R_c\,\widehat{\kappa}_{is}},
\end{equation}
where $R_c = 1.023\cdot 10^6\,\mathrm{chips/s}$ is the spreading-code rate. Given the unjammed carrier-to-noise density ratio $(C_s/N_0)$, the jammer-to-signal ratio $C_i/C_s$, and the estimated SSC, we compute the effective carrier-to-noise density ratio~\citep{kaplan,ruegamer_dissertation},
\begin{equation}
\label{eq:cn0eff_est}
    \left(\frac{C_s}{N_0}\right)_{\mathrm{eff}}
    =
    \frac{\int |H_R(f)|^2 S_s(f)\, df}{
    \frac{1}{(C_s/N_0)} + \left(\frac{C_i}{C_s}\right)\widehat{\kappa}_{is}},
\end{equation}
and report the interference-induced degradation either in linear form,
\begin{equation}
\label{eq:delta_linear}
    \Delta_{\mathrm{lin}} = \left(\frac{C_s}{N_0}\right) - \left(\frac{C_s}{N_0}\right)_{\mathrm{eff}},
\end{equation}
or in the commonly used logarithmic scale (dB-Hz),
\begin{equation}
\label{eq:delta_db}
    \Delta_{\mathrm{dB}} =
    10\log_{10}\!\left(\frac{C_s}{N_0}\right)
    -
    10\log_{10}\!\left(\frac{C_s}{N_0}\right)_{\mathrm{eff}}.
\end{equation}

\textbf{Interference Power Estimation.} For the real-world recordings with the receiver module, $C_i$ is estimated from the recorded IQ samples and the receiver gain setting (VGA) as
\begin{equation}
\label{eq:Ci_est}
    C_i~[\mathrm{dBm}] = P_{\mathrm{sigcon}} - \left(0.5\cdot \mathrm{VGA}-23\right),
\end{equation}
with
\begin{equation}
\label{eq:Psigcon_est}
    P_{\mathrm{sigcon}}
    =
    10\log_{10}\!\Bigg(
    \frac{2\sum_{n=1}^{N} U[n]^2}{N\,R\cdot 1\,\mathrm{mW}}
    \Bigg),
    \qquad
    U[n] = \frac{\sqrt{I[n]^2+Q[n]^2}}{2^8},
\end{equation}
where $R=50\,\Omega$ and the division by $2^8$ accounts for 8-bit quantization. The satellite signal power $C_s$ and noise density $N_0$ are set according to the experimental configuration (real-world open-field vs.~RFCS), yielding all terms required to evaluate Eq.~\ref{eq:cn0eff_est}--\ref{eq:delta_db}~\citep{heublein_benschuh}. We validate the SSC-based impact computation on the Jammertest 2025 dataset, refer to Section~\ref{label_experiments}. We compare the predicted degradation $\Delta$ from the SSC pipeline against measured receiver behavior obtained by post-processing the IQ data with GNSS-SDR, thereby validating the receiver-dependent mapping from spectral overlap and interference power to navigation-relevant $(C_s/N_0)_{\mathrm{eff}}$ loss.

\subsection{Multi-Source Interference Detection}
\label{label_method_multi_source}

\begin{figure*}[!b]
    \centering
	\begin{minipage}[t]{0.495\linewidth}
        \centering
    	\includegraphics[trim=58 126 44 126, clip, width=1.0\linewidth]{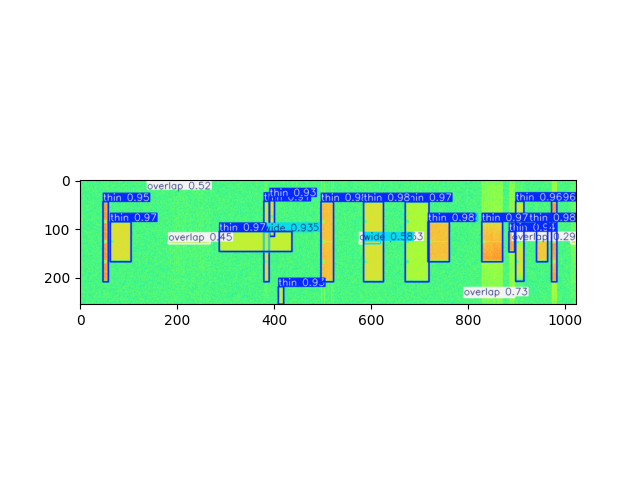}
        \subcaption{YOLOv8s~\citep{wang_yeh_liao}.}
        \label{figure_bounding_box_pretraining1}
    \end{minipage}
    \hfill
	\begin{minipage}[t]{0.495\linewidth}
        \centering
    	\includegraphics[width=1.0\linewidth]{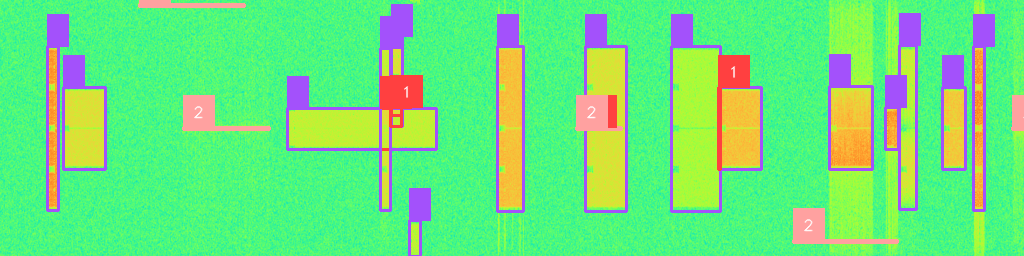}
        \subcaption{RF-DETR~\citep{robinson_robicheaux}.}
        \label{figure_bounding_box_pretraining2}
    \end{minipage}
    \vspace{-0.3cm}
    \caption{Example of multi-source interference detection predictions in GNSS spectrograms. The bounding boxes are annotated with interference-width/overlap classes and confidence scores. The results illustrate the detector's ability to separate narrowband, wideband, and partially overlapping interference events for subsequent per-source characterization.}
    \label{figure_bounding_box_pretraining}
\end{figure*}

\paragraph{Bounding Boxes.} To enable multi-source analysis in realistic spectrograms, we formulate interference localization as an object-detection problem in the time--frequency plane and employ YOLOv8s~\citep{wang_yeh_liao} and RF-DETR~\citep{robinson_robicheaux} as a detector that outputs a set of bounding boxes and confidence scores per spectrogram snapshot. Since dense, manually annotated GNSS spectrograms are scarce, we first pretrain the detector on the Fraunhofer \textit{spectrogram dataset for RF-frame detection} (\href{	http://dx.doi.org/10.24406/fordatis/216}{10.24406/fordatis/216}), proposed by~\cite{wicht_wetzker}, which provides 20{,}000 synthetically generated and augmented signal segments together with bounding-box annotations marking coexisting wireless transmissions (e.g., Wi-Fi and Bluetooth frames). This pretraining step initializes YOLOv8s with transferable features for localizing structured RF energy patterns in time--frequency representations. Figure~\ref{figure_bounding_box_pretraining} visualizes the bounding box predictions of YOLOv8s and RF-DETR on this dataset, visualizing that the prediction is robust for various bandwidths, signal strength, and regarding interference overlapping.

We then transfer the pretrained model to GNSS interference monitoring by applying it to Jammertest spectrograms. For a given spectrogram image $\mathbf{S}\in\mathbb{R}^{F\times T}$, YOLOv8s\footnote{YOLOv8s was initialized from the pretrained \texttt{yolov8s.pt} checkpoint and trained with the Ultralytics framework for 100 epochs using an input image size of $1{,}024 \times 1{,}024$, batch size $4$, and four data-loading workers.} and RF-DETR\footnote{RF-DETR was trained using the \texttt{RFDETRMedium} configuration for 100 epochs on the spectrogram bounding-box dataset. We used an input image size of $1{,}024 \times 1{,}024$, batch size 4, gradient accumulation over 4 steps, and learning rate $10^{-4}$.} predict a set of detections $\mathcal{B}=\big\{(\mathbf{b}_m, s_m)\big\}_{m=1}^{M}$, $\mathbf{b}_m=\big(t_m^{(1)}, f_m^{(1)}, t_m^{(2)}, f_m^{(2)}\big)$, where $\mathbf{b}_m$ denotes the time--frequency bounding box and $s_m$ its confidence. After non-maximum suppression, each box is interpreted as one interferer instance; we crop the corresponding region $\mathbf{S}[\mathbf{b}_m]$ and pass it to the downstream characterization module to obtain per-source estimates of waveform type, occupied bandwidth, and received signal strength. In this way, the overall pipeline decomposes composite scenes into individual interferers, enabling both reliable multi-source detection and subsequent per-interferer characterization and impact analysis.

\paragraph{Bandwidth Estimation from Bounding Boxes.} To further assess whether the object detector provides physically meaningful localization, we evaluate bandwidth estimation directly from the predicted bounding boxes. Since each detected box spans a frequency interval in the spectrogram, its vertical extent can be mapped to an estimated occupied bandwidth. For a detected interference component $m$ with predicted bounding box $\hat{\mathbf{b}}_m$, we convert the lower and upper frequency coordinates of the box into physical frequencies $f_{\min}(\hat{\mathbf{b}}_m)$ and $f_{\max}(\hat{\mathbf{b}}_m)$. The predicted bandwidth is then computed as $\widehat{B}_m = f_{\max}(\hat{\mathbf{b}}_m) - f_{\min}(\hat{\mathbf{b}}_m)$. We compare this estimate with the nominal bandwidth $B_m$ provided by the campaign metadata and report the mean absolute bandwidth error
\begin{equation}
    \mathrm{MAE}_{B}
    =
    \frac{1}{M}
    \sum_{m=1}^{M}
    \left|
    \widehat{B}_m - B_m
    \right|,
\end{equation}
where $M$ denotes the number of matched detected interference components. This metric complements standard detection measures such as confidence and visual localization by quantifying whether the predicted boxes preserve a physically interpretable attribute of the jammer signal. In particular, a low bandwidth error indicates that the detector not only identifies the presence of interference, but also localizes its spectral support accurately enough to support downstream per-source characterization and receiver-aware impact estimation.
\section{DATASET}
\label{label_experiments}

\setlength{\intextsep}{6pt}
\setlength{\columnsep}{12pt}
\begin{wrapfigure}{R}{6.5cm}
    \vspace{-0.25cm}
	\begin{minipage}[t]{0.663\linewidth}
        \centering
    	\includegraphics[width=1.0\linewidth]{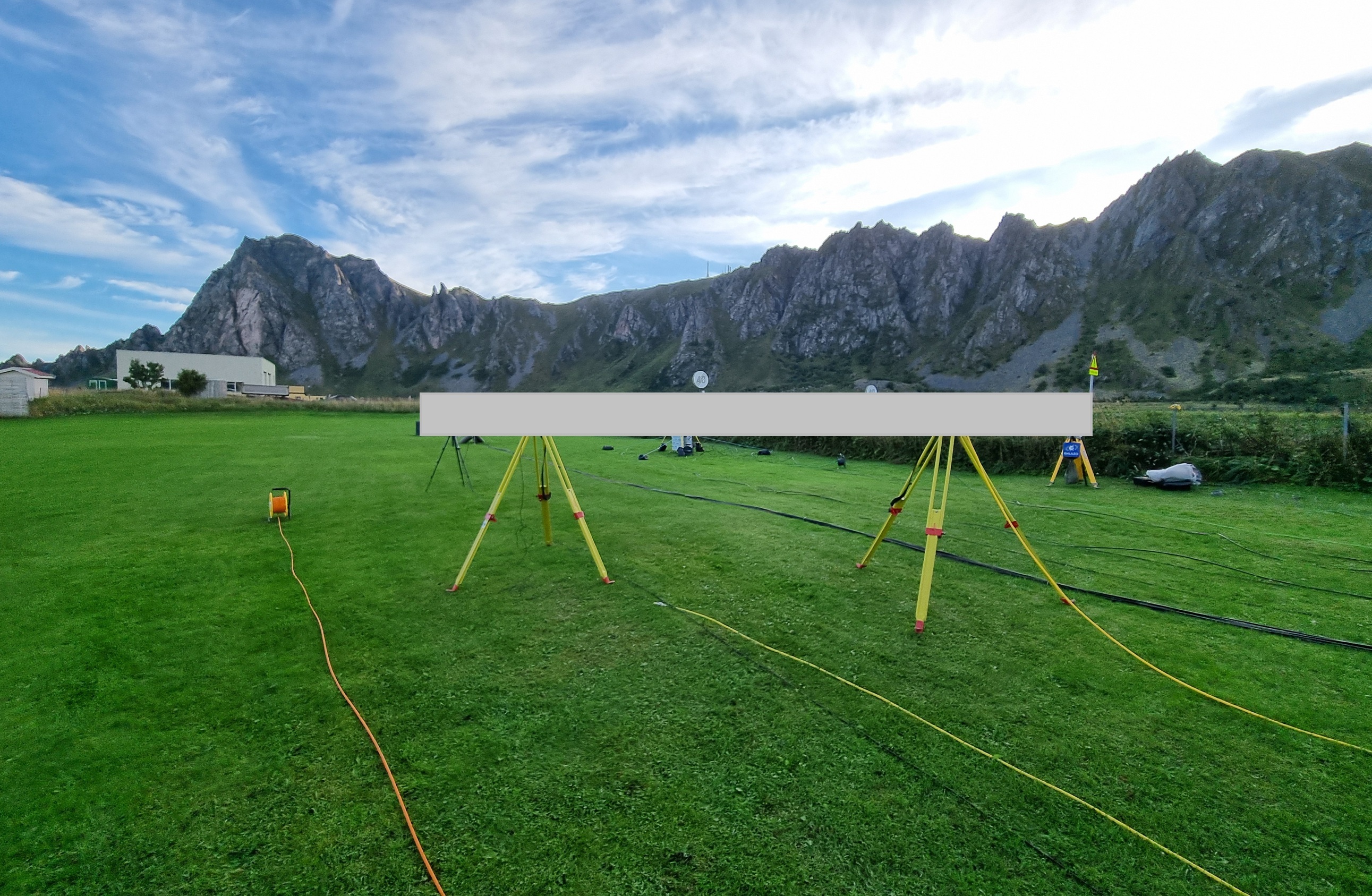}
    \end{minipage}
    \hfill
	\begin{minipage}[t]{0.327\linewidth}
        \centering
    	\includegraphics[width=1.0\linewidth]{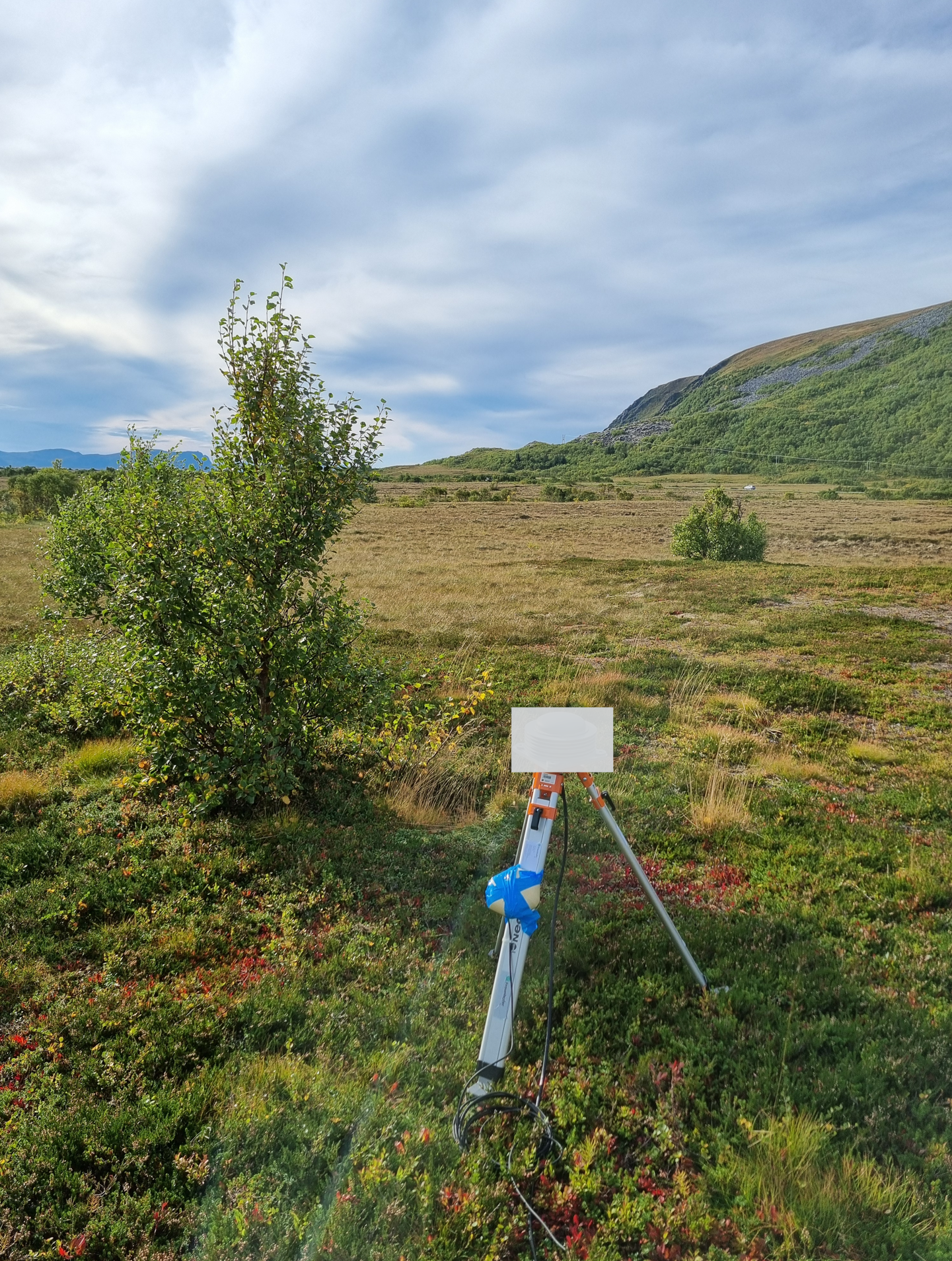}
    \end{minipage}
        \vspace{-0.6cm}
    \caption{Area 1 (left) and Area 2 (right).}
    \label{figure_setup_area}
\end{wrapfigure}

\paragraph{Dataset Setup.} Our dataset was recorded during \texttt{Jammertest\,2025}~\citep{jammertest_gerard}; an authority-controlled, real-world outdoor testbed that enables GNSS jamming, spoofing, and meaconing experiments under safely managed conditions. The campaign infrastructure provides well-defined test cases and a time-resolved transmission plan specifying what is transmitted, where, and when. We focus on the two official outdoor test sites relevant to interference monitoring. \textit{Test Area 1} is centered around the Bleik community house and its nearby surroundings (see Figure~\ref{figure_setup_area}, left). \textit{Test Area 2} is a dedicated open-air setup at a parking lot, structured around surveyed reference points and fixed geometries (see Figure~\ref{figure_setup_area}, right). We participated in \textit{Test Area 1 and Test Area 2} and recorded GNSS data \textit{in parallel} across both sites throughout the campaign schedule. The resulting multi-day, multi-site dataset provides diverse, time-aligned interference conditions under controlled yet realistic outdoor propagation, and it forms the empirical foundation for training and evaluating our ML models for interference analysis and characterization.

\paragraph{Recording Setup.} At both Test Areas 1 and 2, we recorded GNSS data with two antenna/receiver setups operating in parallel to capture complementary views of the interference environment. The first setup is a compact module (\texttt{Innosense}) equipped with a 3G+C antenna and a dual-band E1/E5 receiver front end. It is powered via USB ($5\,\mathrm{V}$, $2\,\mathrm{W}$) and records samples with 8-bit quantization, $50\,\mathrm{MHz}$ bandwidth, and an $81\,\mathrm{MHz}$ sampling rate. The second setup uses a controlled reception pattern antenna (\texttt{CRPA}) consisting of $2 \times 2$-patch elements, capturing $100\,\mathrm{MHz}$ bandwidth data that are quadrature sampled over a $10\,\mu\mathrm{s}$ snapshot duration, enabling array-based interference analysis alongside the single-antenna measurements.

\begin{table*}[!t]
\caption{Test Area~1: unique interference event configurations with aggregated power steps (as listed in the campaign metadata).}
\label{table_intf_area1}
\centering
\scriptsize
\setlength{\tabcolsep}{3.2pt}
\begin{tabular}{clllrrl}
\toprule
\textbf{Day} & \textbf{Jammer ID} & \textbf{Type} & \textbf{Band(s)} & \textbf{BW [MHz]} & \textbf{Power [dBm]} & \textbf{Ref.} \\
\midrule
1 & F8.1 & CW & L1 & 0.1 & 47 & \\
1 & F8.1 & CW & L1,\,G1,\,L2,\,L5,\,E6 & 0.1 & 47 & \\
1 & F8.1 & Sweep & L1 & 20 & 47 & \ref{figure_spectrograms_area1_1} \\
1 & F8.1 & Sweep & L1,\,G1,\,L2,\,L5,\,E6 & 20 & 47 & \\
1 & F8.1 & PRN & E6 & 20 & 47 & \\
1 & F8.1 & PRN & E6,\,E5b & 20 & 47 & \\
1 & F8.1 & PRN & E6,\,E5b,\,L5 & 20 & 47 & \\
1 & F8.1 & PRN & E6,\,E5b,\,L5,\,G2 & 20 & 47 & \\
1 & F8.1 & PRN & E6,\,E5b,\,L5,\,G2,\,L2 & 20 & 47 & \\
1 & F8.1 & PRN & E6,\,E5b,\,L5,\,G2,\,L2,\,B1I & 20 & 47 & \ref{figure_spectrograms_area1_2} \\
1 & F8.1 & PRN & E6,\,E5b,\,L5,\,G2,\,L2,\,B1I,\,G1 & 20 & 47 & \\
1 & F8.1 & PRN & E6,\,E5b,\,L5,\,G2,\,L2,\,B1I,\,G1,\,L1 & 20 & 47 & \ref{figure_spectrograms_area1_14}, \ref{figure_spectrograms_area1_15} \\
1 & F8.1 & PRN & L1 & 10 & 47 & \\
1 & F8.1 & PRN & L1,\,G1,\,L2,\,L5,\,E6 & 10 & 47 & \\
2 & F8.1 & PRN & L1 & 20 & -21 to 47, increments of 2 & \ref{figure_spectrograms_area1_3}, \ref{figure_spectrograms_area1_4}, \ref{figure_spectrograms_area1_5} \\
2 & F8.1 & PRN & L1,\,G1,\,L2,\,L5,\,E6 & 20 & -21 to 47, increments of 2 & \ref{figure_spectrograms_area1_16}\\
2 & F8.1 & PRN & L1,\,L2,\,L5,\,E5a,\,B2a,\,E6,\,G1,\,B1C,\,E1 & 20 & 50 & \\
2 & F1.1 & Meaconing & All & None & 40 & \ref{figure_spectrograms_area1_6} \\
2 & F1.1 & Meaconing & L1,\,E1,\,B1C,\,L2 & None & 40 & \\
2 & F1.1 & Meaconing & L1,\,E1,\,L2,\,B1C & None & 0 to 40, increments of 5 & \ref{figure_spectrograms_area1_7} \\
2 & F1.1 & PRN & L1,\,G1,\,L2,\,L5,\,E6,\,E5b & 20 & 40 & \\
2 & F1.1,\,F8.1 & Meaconing,\,PRN & L5,\,G1,\,G2,\,B1I,\,E5b,\,E6 & 20 & 40 & \ref{figure_spectrograms_area1_8} \\
2 & F8.1 & CW & E6,\,E5b,\,E5a,\,B2a,\,B2b,\,B2I,\,B3I,\,L5,\,G2,\,L2,\,B1I,\,G3 & 0.1 & 47 & \\
2 & F8.1 & CW & L1,\,G1,\,E1,\,B1C,\,B1I & 0.1 & 47 & \\
2 & F1.1 & Meaconing,\,Spoofing & L1,\,E1,\,B1C,\,L2 & None & 40 & \ref{figure_spectrograms_area1_9} \\
2 & M1.1 & PRN & L1,\,L2,\,L5,\,E5a,\,B2a,\,E6,\,G1,\,B1C,\,E1 & 20 & 50 & \\
2 & F8.1,\,M1.1 & PRN & L1,\,L2,\,L5,\,E5a,\,B2a,\,E6,\,G1,\,B1C,\,E1 & 20 & 50 & \\
2 & S & Spoofing & L1,\,E1,\,B1C,\,L2 & None & 40 & \\
3 & S & PRN & L1,\,G1,\,B1I,\,E6,\,L2,\,E5b,\,L5 & 20 & 30 & \ref{figure_spectrograms_area1_11} \\
3 & S & PRN & L1,\,L2,\,L5,\,E1,\,E5a,\,E5b,\,E6 & 20 & 30 & \\
3 & S & Spoofing & E1 & None & 25 & \\
3 & S & Spoofing & E1,\,E5a,\,E5b,\,E6 & None & 25 & \\
3 & S & Spoofing & L1 & None & 25 & \ref{figure_spectrograms_area1_10} \\
3 & S & Spoofing & L1,\,E1 & None & 25 & \\
3 & S & Spoofing & L1,\,E1,\,E6,\,E5a,\,L2,\,E5b,\,L5 & None & 25 & \\
3 & S & Spoofing & L1,\,L2,\,L5 & None & 25 & \\
3 & S & Spoofing & L1,\,L2,\,L5,\,E1,\,E5a,\,E5b,\,E6 & None & -35 to 25, increments of 5 & \ref{figure_spectrograms_area1_12} \\
4 & S & PRN & L1,\,L2,\,L5,\,E1,\,E5a,\,E5b,\,E6 & 20 & 30 & \\
4 & S & Spoofing & E1 & None & 15 & \\
4 & S & Spoofing & L1 & None & 15 & \\
4 & S & Spoofing & L1,\,E1 & None & -20 & \\
4 & S & Spoofing & L1,\,L2,\,L5,\,E1,\,E5a,\,E5b,\,E6 & None & -35 to 15, increments of 5 & \ref{figure_spectrograms_area1_17}, \ref{figure_spectrograms_area1_18} \\
5 & F1.1 & Meaconing,\,Spoofing & L1,\,E1,\,B1C,\,L2 & None & 40 & \ref{figure_spectrograms_area1_13} \\
\bottomrule
\end{tabular}%
\end{table*}

\newcommand\x{0.105}
\begin{figure*}[!t]
\captionsetup[subfigure]{font=scriptsize}
    \centering
	\begin{minipage}[t]{\x\linewidth}
        \centering
    	\includegraphics[trim=12 6 26 24, clip, width=1.0\linewidth]{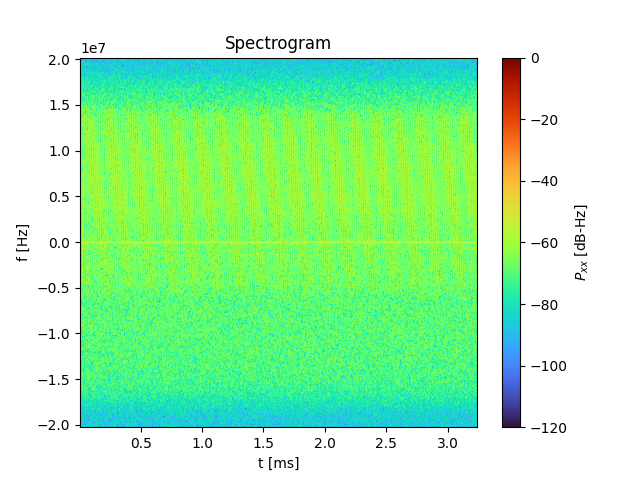}
        \subcaption{Area 1, E1, F8.1, Sweep, $\text{BW}=20\,\text{MHz}$.}
        \label{figure_spectrograms_area1_1}
    \end{minipage}
    \hfill
	\begin{minipage}[t]{\x\linewidth}
        \centering
    	\includegraphics[trim=12 6 26 24, clip, width=1.0\linewidth]{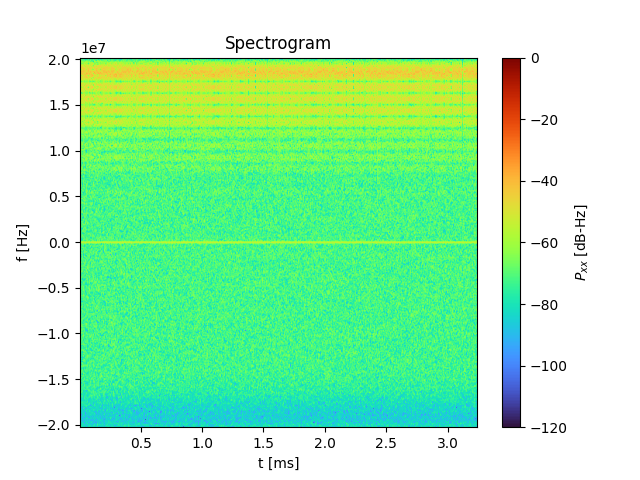}
        \subcaption{Area 1, E1, F8.1, PRN, $\text{BW}=20\,\text{MHz}$.}
        \label{figure_spectrograms_area1_2}
    \end{minipage}
    \hfill
	\begin{minipage}[t]{\x\linewidth}
        \centering
    	\includegraphics[trim=12 6 26 24, clip, width=1.0\linewidth]{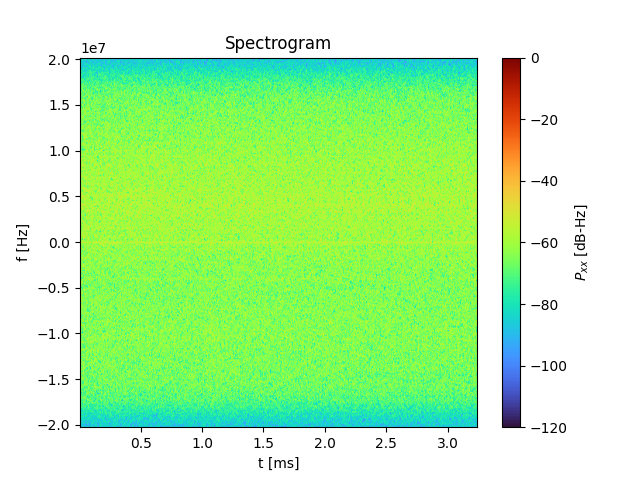}
        \subcaption{Area 1, E1, F8.1, PRN, $-3\,\text{dBm}$.}
        \label{figure_spectrograms_area1_3}
    \end{minipage}
    \hfill
	\begin{minipage}[t]{\x\linewidth}
        \centering
    	\includegraphics[trim=12 6 26 24, clip, width=1.0\linewidth]{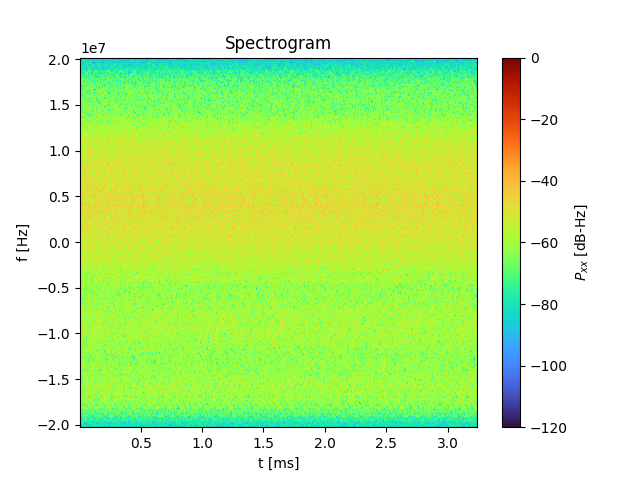}
        \subcaption{Area 1, E1, F8.1, PRN, $9\,\text{dBm}$.}
        \label{figure_spectrograms_area1_4}
    \end{minipage}
    \hfill
	\begin{minipage}[t]{\x\linewidth}
        \centering
    	\includegraphics[trim=12 6 26 24, clip, width=1.0\linewidth]{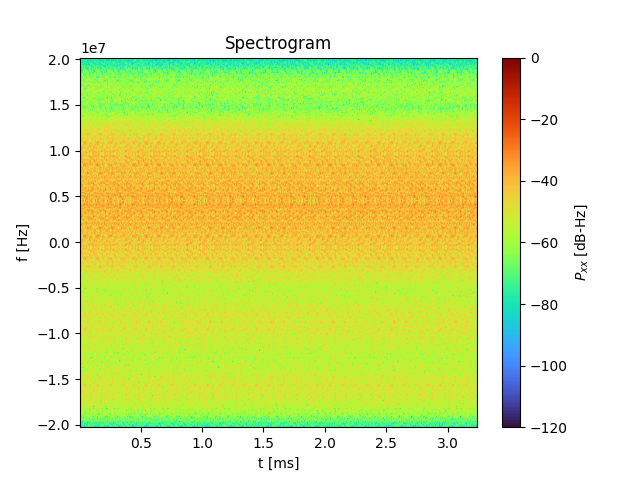}
        \subcaption{Area 1, E1, F8.1, PRN, $19\,\text{dBm}$.}
        \label{figure_spectrograms_area1_5}
    \end{minipage}
    \hfill
	\begin{minipage}[t]{\x\linewidth}
        \centering
    	\includegraphics[trim=12 6 26 24, clip, width=1.0\linewidth]{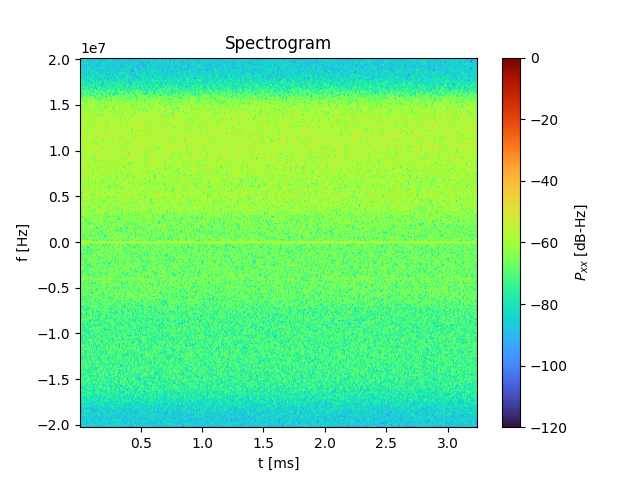}
        \subcaption{Area 1, E1, F1.1, Meaconing, $40\,\text{dBm}$.}
        \label{figure_spectrograms_area1_6}
    \end{minipage}
    \hfill
	\begin{minipage}[t]{\x\linewidth}
        \centering
    	\includegraphics[trim=12 6 26 24, clip, width=1.0\linewidth]{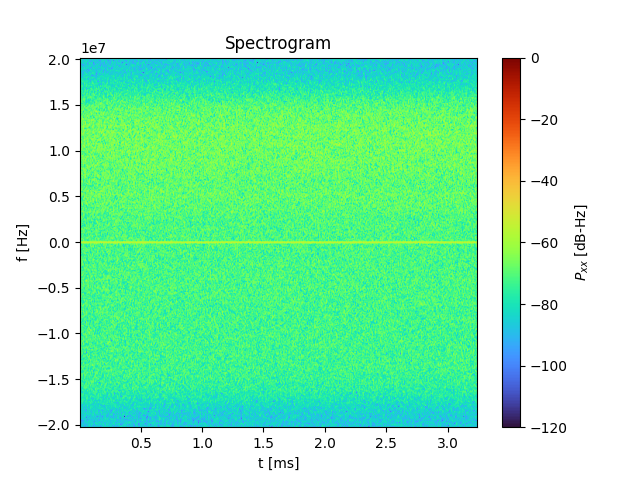}
        \subcaption{Area 1, E1, F1.1, Meaconing, $30\,\text{dBm}$.}
        \label{figure_spectrograms_area1_7}
    \end{minipage}
    \hfill
	\begin{minipage}[t]{\x\linewidth}
        \centering
    	\includegraphics[trim=12 6 26 24, clip, width=1.0\linewidth]{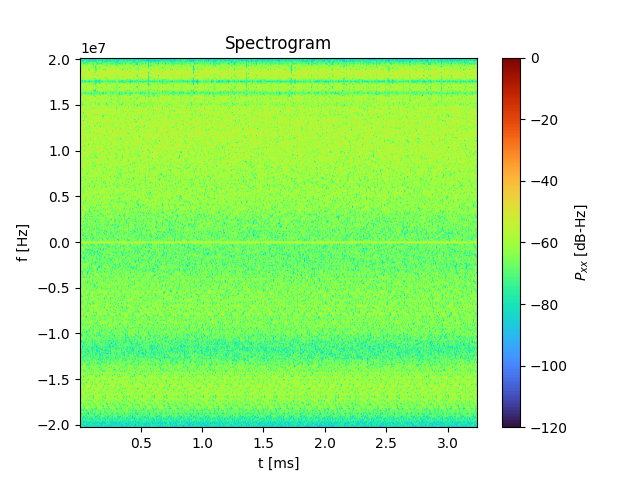}
        \subcaption{Area 1, E1, F1.1, F8.1, Meaconing+PRN.}
        \label{figure_spectrograms_area1_8}
    \end{minipage}
    \hfill
	\begin{minipage}[t]{\x\linewidth}
        \centering
    	\includegraphics[trim=12 6 26 24, clip, width=1.0\linewidth]{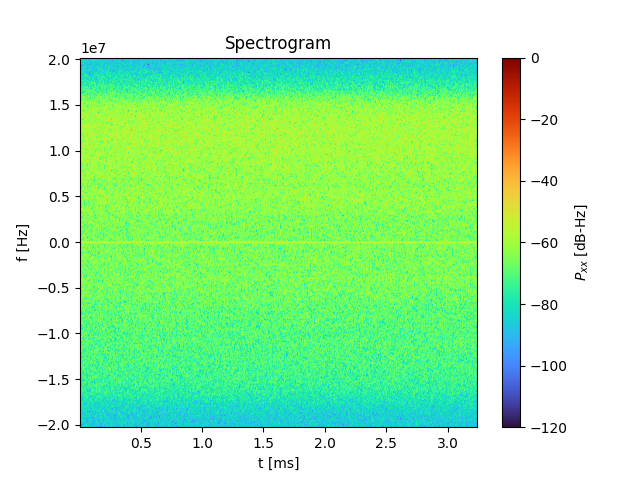}
        \subcaption{Area 1, E1, F1.1, F8.1, Meac+Spoof.}
        \label{figure_spectrograms_area1_9}
    \end{minipage}
	\begin{minipage}[t]{\x\linewidth}
        \centering
    	\includegraphics[trim=12 6 26 24, clip, width=1.0\linewidth]{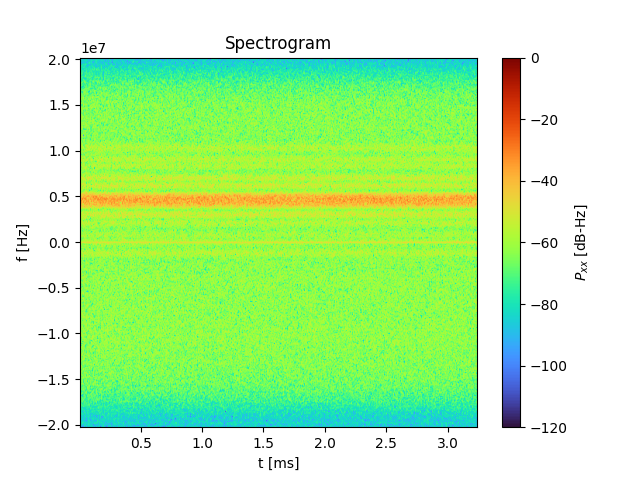}
        \subcaption{Area 1, E1, S, Spoofing, $25\,\text{dBm}$.}
        \label{figure_spectrograms_area1_10}
    \end{minipage}
    \hfill
	\begin{minipage}[t]{\x\linewidth}
        \centering
    	\includegraphics[trim=12 6 26 24, clip, width=1.0\linewidth]{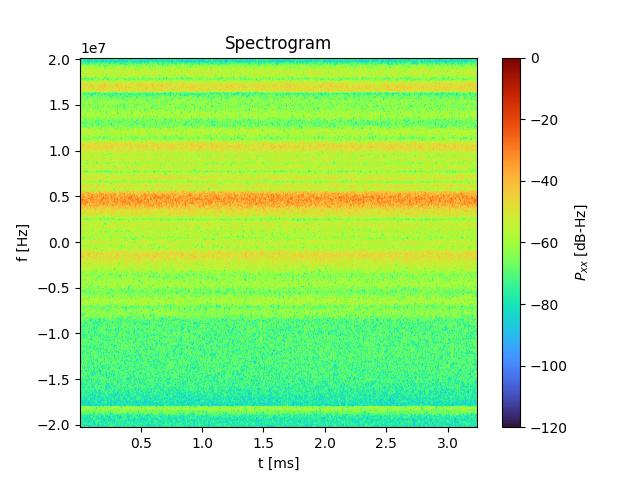}
        \subcaption{Area 1, E1, S, PRN, $30\,\text{dBm}$.}
        \label{figure_spectrograms_area1_11}
    \end{minipage}
    \hfill
	\begin{minipage}[t]{\x\linewidth}
        \centering
    	\includegraphics[trim=12 6 26 24, clip, width=1.0\linewidth]{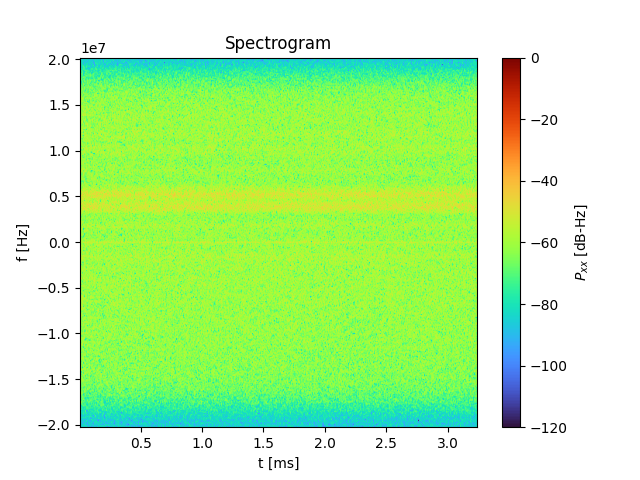}
        \subcaption{Area 1, E1, S, Spoofing, $-5\,\text{dBm}$.}
        \label{figure_spectrograms_area1_12}
    \end{minipage}
    \hfill
	\begin{minipage}[t]{\x\linewidth}
        \centering
    	\includegraphics[trim=12 6 26 24, clip, width=1.0\linewidth]{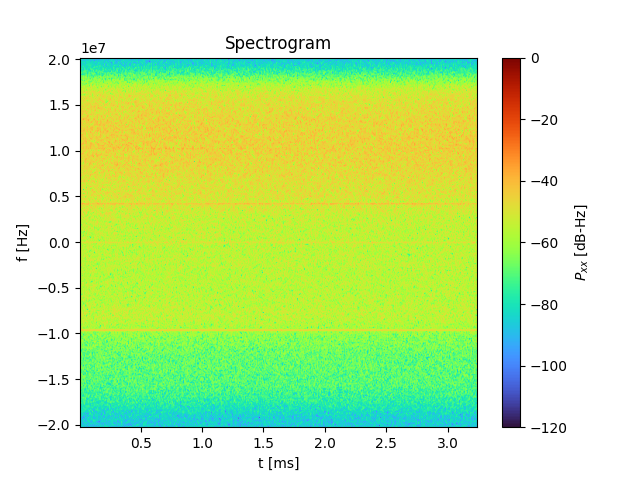}
        \subcaption{Area 1, E1, F1.1, $40\,\text{dBm}$, Meac+Spoof.}
        \label{figure_spectrograms_area1_13}
    \end{minipage}
    \hfill
	\begin{minipage}[t]{\x\linewidth}
        \centering
    	\includegraphics[trim=12 6 26 24, clip, width=1.0\linewidth]{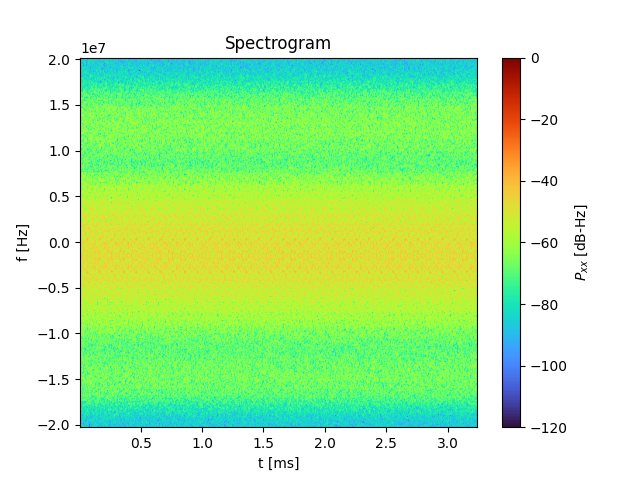}
        \subcaption{Area 1, E5, F8.1, PRN, $47\,\text{dBm}$.}
        \label{figure_spectrograms_area1_14}
    \end{minipage}
    \hfill
	\begin{minipage}[t]{\x\linewidth}
        \centering
    	\includegraphics[trim=12 6 26 24, clip, width=1.0\linewidth]{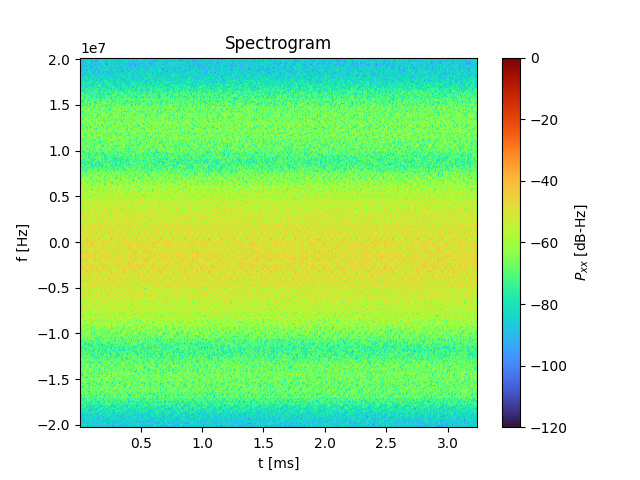}
        \subcaption{Area 1, E5, F8.1, PRN, $47\,\text{dBm}$.}
        \label{figure_spectrograms_area1_15}
    \end{minipage}
    \hfill
	\begin{minipage}[t]{\x\linewidth}
        \centering
    	\includegraphics[trim=12 6 26 24, clip, width=1.0\linewidth]{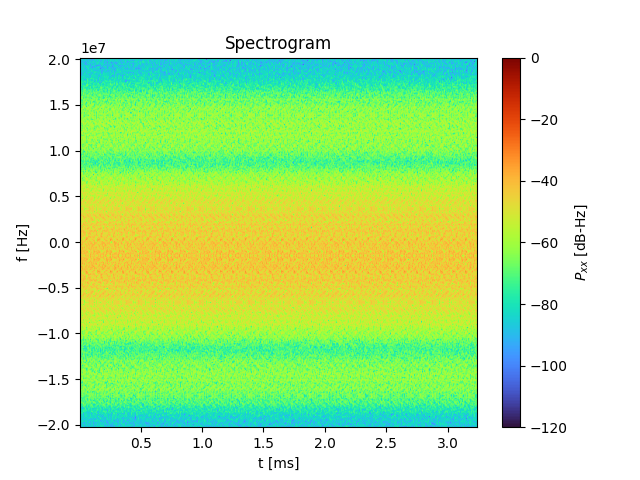}
        \subcaption{Area 1, E5, F8.1, PRN, $9\,\text{dBm}$.}
        \label{figure_spectrograms_area1_16}
    \end{minipage}
    \hfill
	\begin{minipage}[t]{\x\linewidth}
        \centering
    	\includegraphics[trim=12 6 26 24, clip, width=1.0\linewidth]{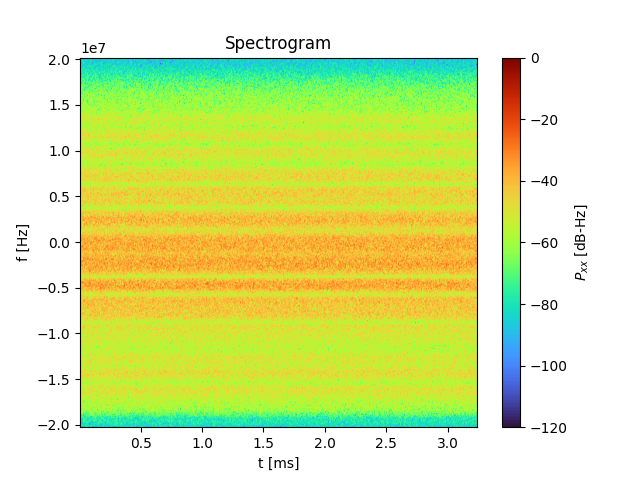}
        \subcaption{Area 1, E5, S, Spoofing, $10\,\text{dBm}$.}
        \label{figure_spectrograms_area1_17}
    \end{minipage}
    \hfill
	\begin{minipage}[t]{\x\linewidth}
        \centering
    	\includegraphics[trim=12 6 26 24, clip, width=1.0\linewidth]{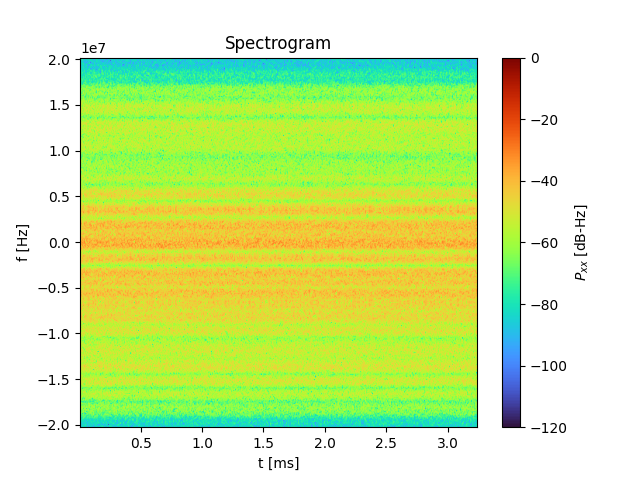}
        \subcaption{Area1, E5, S, PRN, $30\,\text{dBm}$.}
        \label{figure_spectrograms_area1_18}
    \end{minipage}
    \vspace{-0.3cm}
    \caption{Exemplary jamming and spoofing events in Area 1 for both bands E1 and E5.}
    \label{figure_spectrograms_area1}
\end{figure*}

\begin{table*}[!t]
\caption{Test Area~2: unique interference event configurations with aggregated power steps (as listed in the campaign metadata).}
\label{table_intf_area2}
\vspace{-0.2cm}
\centering
\scriptsize
\setlength{\tabcolsep}{2.2pt}
\begin{tabular}{clllrrl}
\toprule
\textbf{Day} & \textbf{Jammer ID} & \textbf{Type} & \textbf{Band(s)} & \textbf{BW} & \textbf{Power} & \textbf{Ref.} \\
\midrule
1 & S1.1 & Chirp & L1,\,E1,\,B1C,\,B1I & 30 & 15 & \\
1 & S2.1 & ChirpB & E6,\,E5b,\,E5a,\,B2a,\,B2I,\,B2b,\,G2,\,G3,\,L2,\,L5,\,B1I,\,B1C,\,E1,\,L1,\,G1 & 85 & 20 & \\
1 & U1.1 & Chirp & L1,\,G1,\,E1,\,B1C,\,B1I & 80 & -50 & \\
1 & H1.1 & Mod & L1,\,E1,\,L2,\,B1C & 20 & 20 & \\
1 & H1.2 & Chirp & L1,\,E1,\,B1C & 20 & 15 & \\
1 & H3.1 & Chirp & L1,\,E1,\,B1C & 28 & 20 & \\
1 & H3.3 & Chirp & L1,\,L2,\,E1,\,B1C,\,L5,\,E5a,\,B2a & 20 & 30 & \\
1 & H4.1 & ChirpM & E6,\,E5b,\,E5a,\,B2a,\,B2I,\,B2b,\,G2,\,G3,\,L2,\,L5,\,B3I,\,B1I,\,B1C,\,E1,\,L1,\,G1 & 102 & 25 & \\
1 & H6.1 & Chirp & L1,\,E1,\,G1,\,B1C & 87 & 25 & \\
1 & H6.2 & ChirpMS & E6,\,E5b,\,E5a,\,B2a,\,B2I,\,B2b,\,G2,\,G3,\,L2,\,L5,\,B3I,\,B1C,\,E1,\,L1 & 30 & 30 & \\
1 & H6.3 & Chirp & E6,\,E5b,\,E5a,\,B2a,\,B2I,\,B2b,\,G2,\,G3,\,L2,\,L5,\,B3I,\,B1C,\,E1,\,L1 & 26 & 30 & \\
1 & H6.4 & Triang & E6,\,E5b,\,E5a,\,B2a,\,B2I,\,B2b,\,G2,\,G3,\,L2,\,L5,\,B3I,\,B1I,\,B1C,\,E1,\,L1,\,G1 & 81 & 30 & \\
1 & H8.1 & Triang & L1,\,E1,\,G1,\,B1C,\,B1I & 77 & 25 & \\
1 & F6.1 & FmS & E6,\,E5b,\,E5a,\,B2a,\,B2I,\,B2b,\,G2,\,G3,\,L2,\,L5,\,B3I,\,B1I,\,B1C,\,E1,\,L1,\,G1 & 70 & 35 & \\
2 & S1.1,\,S1.2,\,S1.3 & Chirp & G1,\,L1,\,E1,\,B1C,\,B1I & 30 & 20 & \ref{figure_spectrograms_area2_1} \\
2 & S2.1,\,S2.2,\,S2.3 & Chirp & G1,\,L1,\,E1,\,B1C,\,B1I & 30 & 30 & \\
2 & S2.1,\,S2.2,\,S2.3 & ChirpB & G1,\,L1,\,E1,\,B1C,\,B1I & 85 & 30 & \ref{figure_spectrograms_area2_2} \\
2 & U1.1,\,U1.2,\,U1.3 & Chirp & G1,\,L1,\,E1,\,B1C,\,B1I & 80 & -50 & \ref{figure_spectrograms_area2_3} \\
2 & H6.4,\,H6.5,\,H6.6 & Triang & G1,\,L1,\,E1,\,B1C,\,B1I & 80 & 30 & \ref{figure_spectrograms_area2_4} \\
2 & H6.4,\,H6.5,\,H6.6 & Triang & G1,\,L1,\,E1,\,B1C,\,B1I & None & 30 & \\
2 & H1.1,\,H1.4,\,H1.5 & Chirp & L1,\,E1,\,B1C,\,L2 & 20 & 20 & \\
2 & H1.1,\,H1.4,\,H1.5 & PRN & L1,\,E1,\,B1C,\,L2 & 2 & 35 & \ref{figure_spectrograms_area2_5} \\
2 & H1.1,\,H1.4,\,H1.5 & PRN & L1,\,E1,\,B1C,\,L2 & 20 & 35 & \\
2 & H1.1,\,H1.4,\,H1.5 & CW & L1,\,E1,\,B1C,\,L2 & 0.1 & 35 & \ref{figure_spectrograms_area2_6} \\
3 & M1.1 & Spoofing & L1,\,E1 & None & 0,\,5,\,10 & \\
3 & M1.1,\,H1.1,\,H1.4,\,H1.5 & Chirp & L1,\,L2,\,E1 & 20 & 35 & \ref{figure_spectrograms_area2_7} \\
3 & M1.1,\,H1.1,\,H1.4,\,H1.5 & Chirp,\,Spoofing & L1,\,L2,\,E1 & 20 & 35 & \\
3 & M1.1,\,H1.1,\,H1.4,\,H1.5 & Spoofing & L1,\,L2,\,E1 & 20 & 5 & \\
3 & M1.1,\,H1.1,\,H1.4,\,H1.5,\,H1.6,\,H1.7 & Chirp,\,PRN & L1,\,L2,\,E1 & 20 & 35 & \ref{figure_spectrograms_area2_8} \\
3 & M1.1,\,H1.1,\,H1.4,\,H1.5,\,H1.6,\,H1.7,\,H6.5,\,H6.6 & Chirp,\,PRN,\,Triang & L1,\,L2,\,E1 & 80 & 35 & \\
4 & S1.1 & Chirp & L1 & 30 & 25 & \ref{figure_spectrograms_area2_9} \\
\bottomrule
\end{tabular}%
\end{table*}

\begin{figure*}[!t]
\captionsetup[subfigure]{font=scriptsize}
    \centering
	\begin{minipage}[t]{\x\linewidth}
        \centering
    	\includegraphics[trim=12 6 26 24, clip, width=1.0\linewidth]{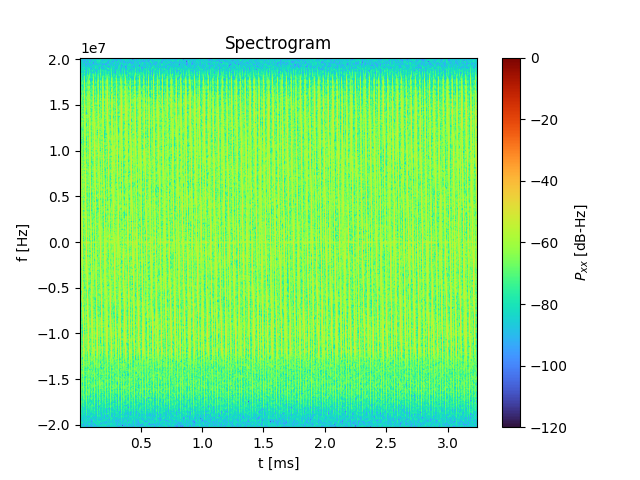}
        \subcaption{Area2, E1, S1.1,\,S1.2,\,S1.3, Chirp, $\text{BW}=30\,\text{MHz}$, $20\,\text{dBm}$.}
        \label{figure_spectrograms_area2_1}
    \end{minipage}
    \hfill
	\begin{minipage}[t]{\x\linewidth}
        \centering
    	\includegraphics[trim=12 6 26 24, clip, width=1.0\linewidth]{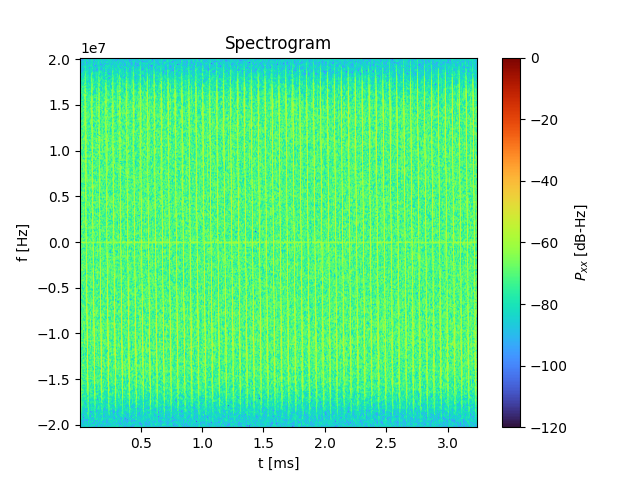}
        \subcaption{Area2, E1, S2.1,\,S2.2,\,S2.3, ChirpB, $\text{BW}=85\,\text{MHz}$, $30\,\text{dBm}$.}
        \label{figure_spectrograms_area2_2}
    \end{minipage}
    \hfill
	\begin{minipage}[t]{\x\linewidth}
        \centering
    	\includegraphics[trim=12 6 26 24, clip, width=1.0\linewidth]{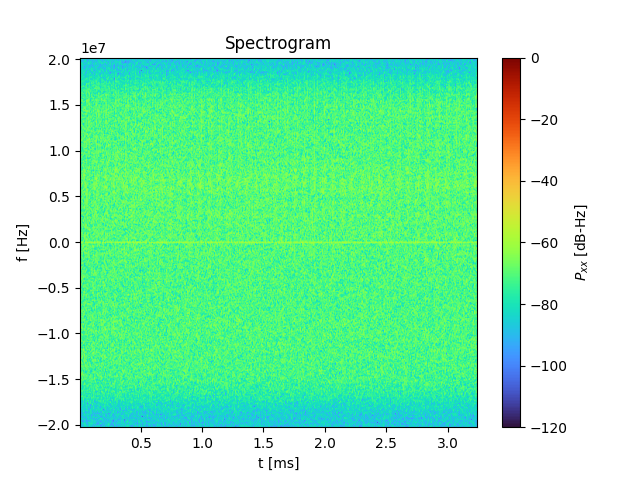}
        \subcaption{Area2, E1, U1.1,\,U1.2,\,U1.3, Chirp, $\text{BW}=80\,\text{MHz}$, $-50\,\text{dBm}$.}
        \label{figure_spectrograms_area2_3}
    \end{minipage}
    \hfill
	\begin{minipage}[t]{\x\linewidth}
        \centering
    	\includegraphics[trim=12 6 26 24, clip, width=1.0\linewidth]{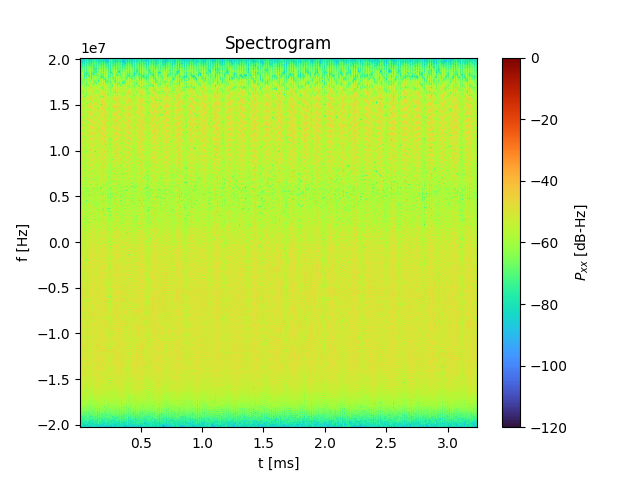}
        \subcaption{Area2, E1, H6.4,\,H6.5,\,H6.6, Triang, $\text{BW}=80\,\text{MHz}$, $30\,\text{dBm}$.}
        \label{figure_spectrograms_area2_4}
    \end{minipage}
    \hfill
	\begin{minipage}[t]{\x\linewidth}
        \centering
    	\includegraphics[trim=12 6 26 24, clip, width=1.0\linewidth]{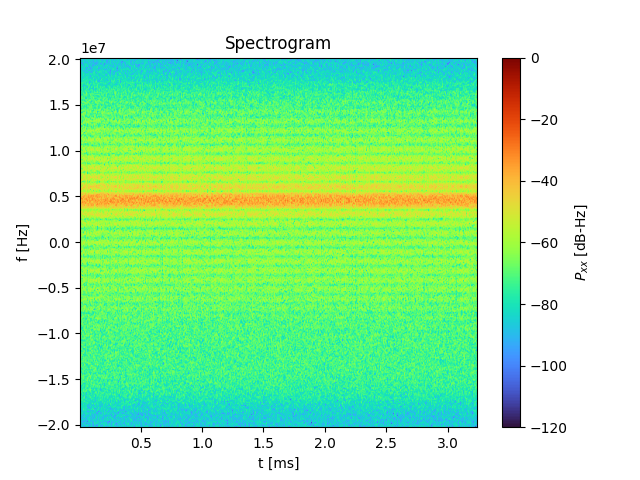}
        \subcaption{Area2, E1, H1.1,\,H1.4,\,H1.5, PRN, $\text{BW}=2\,\text{MHz}$, $35\,\text{dBm}$.}
        \label{figure_spectrograms_area2_5}
    \end{minipage}
    \hfill
	\begin{minipage}[t]{\x\linewidth}
        \centering
    	\includegraphics[trim=12 6 26 24, clip, width=1.0\linewidth]{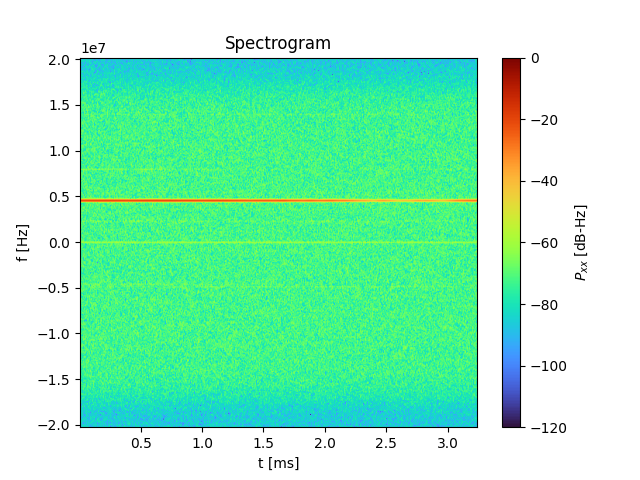}
        \subcaption{Area2, E1, H1.1,\,H1.4,\,H1.5, CW, $\text{BW}=0.1\,\text{MHz}$, $35\,\text{dBm}$.}
        \label{figure_spectrograms_area2_6}
    \end{minipage}
    \hfill
	\begin{minipage}[t]{\x\linewidth}
        \centering
    	\includegraphics[trim=12 6 26 24, clip, width=1.0\linewidth]{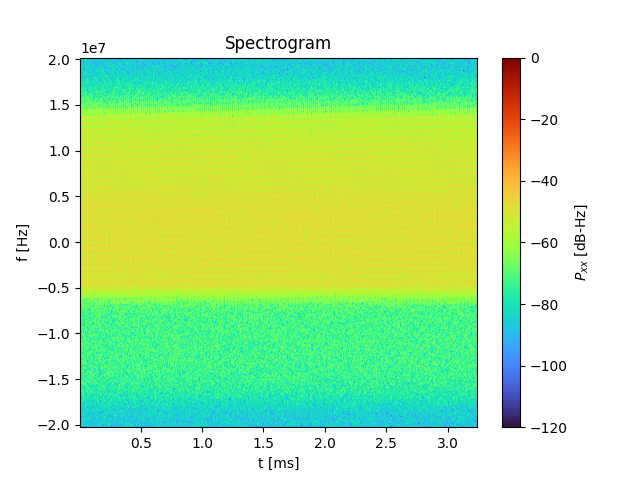}
        \subcaption{Area2, E1, M1.1,\,H1.1,\,H1.4, H1.5, Chirp, $\text{BW}=20\,\text{MHz}$, $35\,\text{dBm}$.}
        \label{figure_spectrograms_area2_7}
    \end{minipage}
    \hfill
	\begin{minipage}[t]{\x\linewidth}
        \centering
    	\includegraphics[trim=12 6 26 24, clip, width=1.0\linewidth]{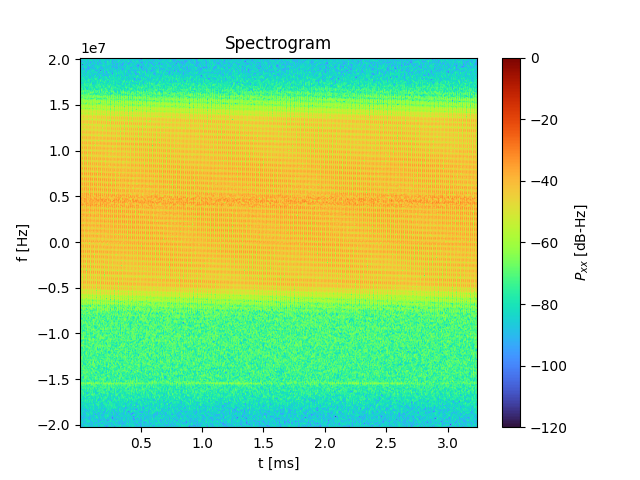}
        \subcaption{Area2, E1, M1.1,\,H1.1,\,H1.4, H1.5,\,H1.6,\,H1.7, Chirp+PRN, $\text{BW}=20\,\text{MHz}$, $35\,\text{dBm}$.}
        \label{figure_spectrograms_area2_8}
    \end{minipage}
    \hfill
	\begin{minipage}[t]{\x\linewidth}
        \centering
    	\includegraphics[trim=12 6 26 24, clip, width=1.0\linewidth]{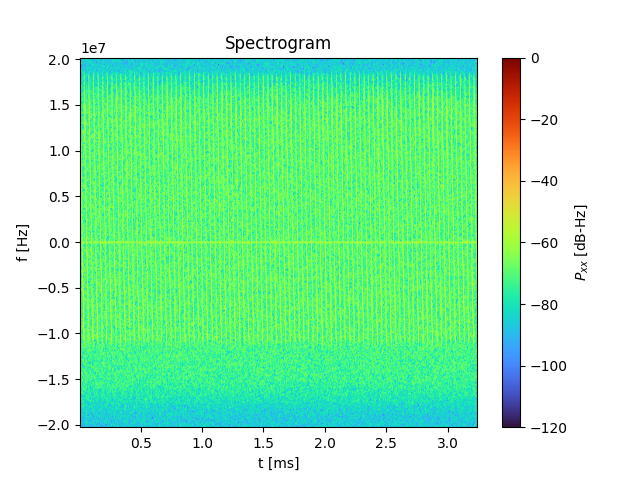}
        \subcaption{Area2, E1, S1.1, Chirp, $\text{BW}=20\,\text{MHz}$, $35\,\text{dBm}$.}
        \label{figure_spectrograms_area2_9}
    \end{minipage}
    \vspace{-0.3cm}
    \caption{Exemplary jamming and spoofing events in Area 2 for band E1.}
    \label{figure_spectrograms_area2}
\end{figure*}

\paragraph{Dataset Details.} The dataset is accompanied by per-recording metadata (e.g., start/end times, configured bandwidth, and signal type), which enables the assignment of time-aligned ground-truth labels for supervised learning and benchmarking\footnote{Jammertest 2025 transmission plan, see github: \href{https://github.com/NPRA/jammertest-plan}{https://github.com/NPRA/jammertest-plan}}. Tables~\ref{table_intf_area1}--\ref{table_intf_area2} give an overview of all recorded interference events. In Area 1, the recorded interference scenarios include continuous-wave (CW) emissions with very narrow bandwidths (e.g., $\approx 0.1\,\text{MHz}$), frequency-sweep signals (e.g., $\approx 20\,\text{MHz}$), and pseudo-random noise (PRN)-modulated jamming (typically $\approx 10$--$20\,\text{MHz}$), transmitted on single bands (e.g., L1/E1) as well as multi-band configurations spanning GPS/Galileo/BeiDou components. The dataset also comprises meaconing and spoofing runs, including mixed sequences (e.g., Meac+PRN and Meac+Spoof) and dedicated power-variation experiments that intentionally change interference strength to capture realistic variability. In Area 2, the signal catalogue is broader and features several chirp/sweep families (e.g., Chirp, ChirpB, ChirpM, ChirpMS) with bandwidths from roughly $20\,\text{MHz}$ up to $\approx 100\,\text{MHz}$, as well as additional modulation classes such as Triang ($\approx 77$--$81\,\text{MHz}$), FmS (FM-sweep, $\approx 70\,\text{MHz}$), and Mod ($\approx 20\,\text{MHz}$), complemented by PRN and CW baseline recordings. Moreover, Area 2 includes multi-emitter configurations and spoofing intervals across single- and multi-band settings. Figures~\ref{figure_spectrograms_area1}--\ref{figure_spectrograms_area2} visualize exemplary jamming and spoofing events.

\newcommand\y{0.16}
\begin{figure*}[!t]
\captionsetup[subfigure]{font=scriptsize}
    \centering
	\begin{minipage}[t]{\y\linewidth}
        \centering
    	\includegraphics[trim=11 10 10 10, clip, width=1.0\linewidth]{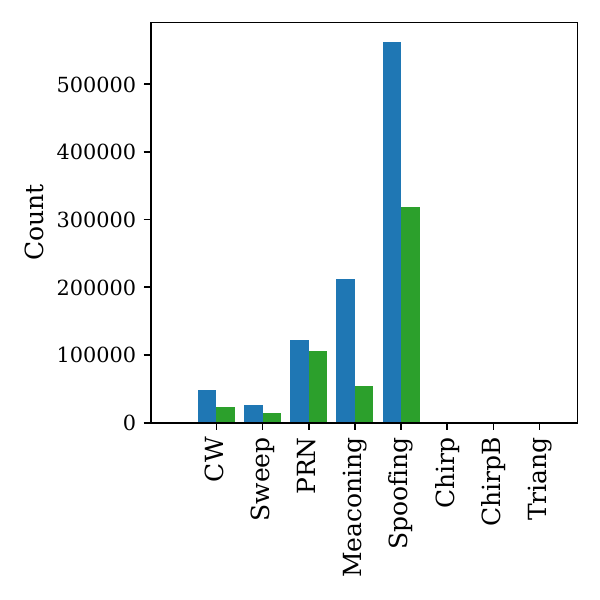}
        \subcaption{Innosense, Area 1.}
        \label{figure_dataset_stats1}
    \end{minipage}
    \hfill
	\begin{minipage}[t]{\y\linewidth}
        \centering
    	\includegraphics[trim=11 10 10 10, clip, width=1.0\linewidth]{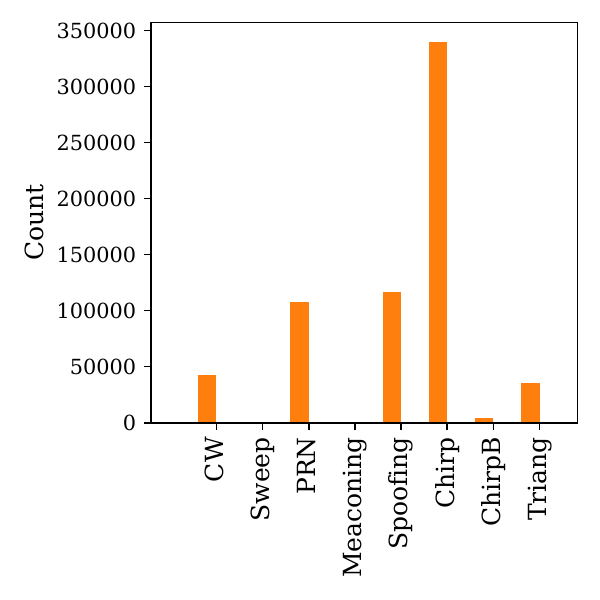}
        \subcaption{Innosense, Area 2.}
        \label{figure_dataset_stats2}
    \end{minipage}
    \hfill
	\begin{minipage}[t]{\y\linewidth}
        \centering
    	\includegraphics[trim=11 10 10 10, clip, width=1.0\linewidth]{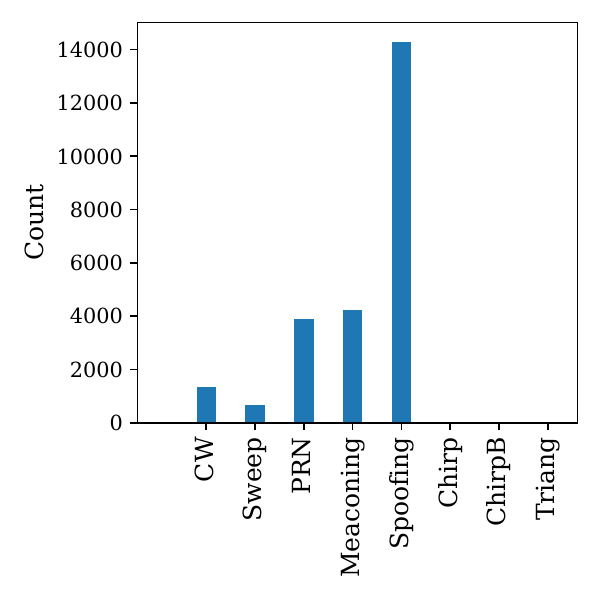}
        \subcaption{CRPA, Area 1.}
        \label{figure_dataset_stats3}
    \end{minipage}
    \hfill
	\begin{minipage}[t]{\y\linewidth}
        \centering
    	\includegraphics[trim=11 10 10 10, clip, width=1.0\linewidth]{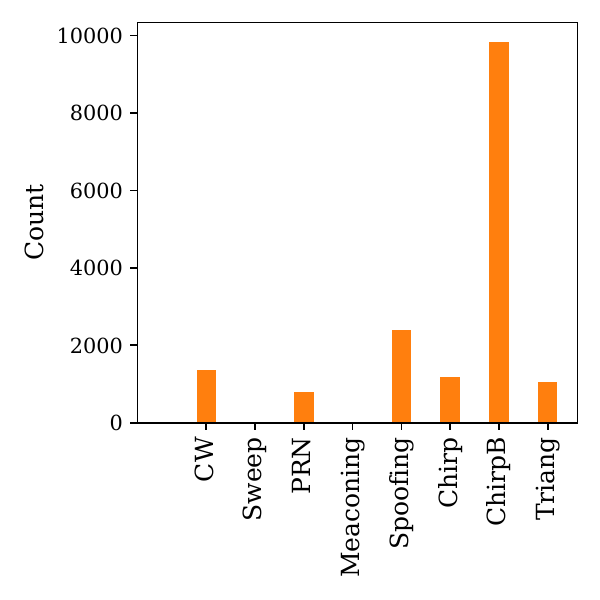}
        \subcaption{CRPA, Area 2.}
        \label{figure_dataset_stats4}
    \end{minipage}
    \hfill
	\begin{minipage}[t]{\y\linewidth}
        \centering
    	\includegraphics[trim=11 10 10 10, clip, width=1.0\linewidth]{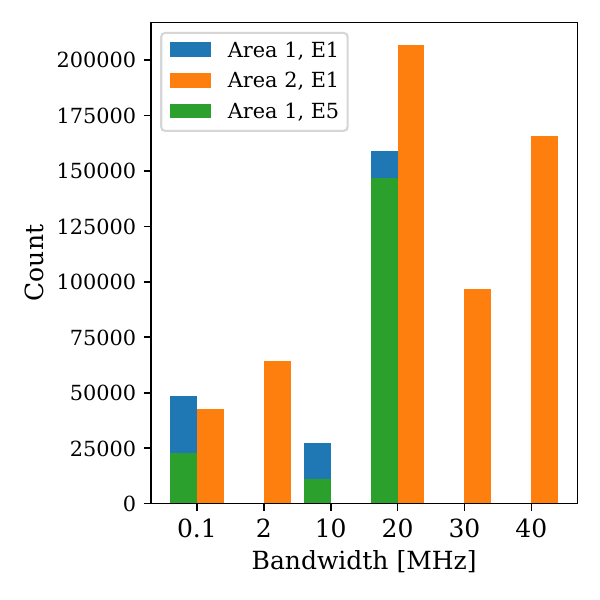}
        \subcaption{Histogram of bandwidths for Innosense.}
        \label{figure_dataset_stats5}
    \end{minipage}
    \hfill
	\begin{minipage}[t]{\y\linewidth}
        \centering
    	\includegraphics[trim=11 10 10 10, clip, width=1.0\linewidth]{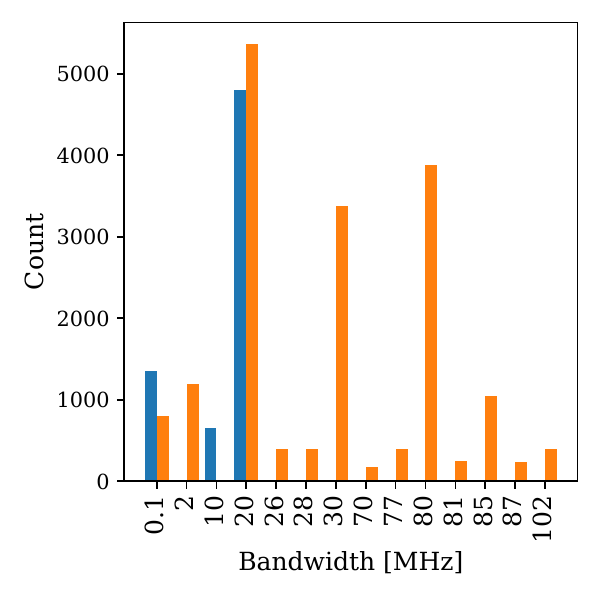}
        \subcaption{Histogram of bandwidths for CRPA.}
        \label{figure_dataset_stats6}
    \end{minipage}
    \vspace{-0.2cm}
    \caption{Overview of number of samples for each area, receiver, band E1/E5, and bandwidth.}
    \label{figure_dataset_stats}
\end{figure*}

\paragraph{Dataset Statistics.} Figure~\ref{figure_dataset_stats} summarizes the class and bandwidth distributions of the published dataset across receiver setups and test areas. Figures~\ref{figure_dataset_stats1}--\ref{figure_dataset_stats4} report the number of labeled samples per interference family (CW, Sweep, PRN, Meaconing, Spoofing, Chirp, ChirpB, Triang) for Innosense and CRPA in Area~1 and Area~2, respectively, and show that both areas contain a broad mix of jamming and deception scenarios, with noticeable imbalances across classes and differences between areas that reflect the campaign schedule. Figures~\ref{figure_dataset_stats5}--\ref{figure_dataset_stats6} visualize the discrete bandwidth configurations observed in the dataset: Innosense concentrates on a small set of nominal bandwidths (e.g., $0.1$\,/\,$2$\,/\,$10$\,/\,$20$\,/\,$30$\,/\,$40\,\text{MHz}$) with separate contributions from Area~1 (E1 and E5) and Area~2 (E1), consistent with the receiver’s limited effective bandwidth. In contrast, CRPA spans a wider range of bandwidth settings, including wideband configurations (up to $\approx 100\,\text{MHz}$-class events) and additional intermediate bandwidths, with different occupancy between Area~1 and Area~2, highlighting that the wideband array recordings capture interference regimes that are underrepresented in the single-antenna setup. Overall, these statistics confirm that the dataset is heterogeneous across (receiver, area, band, bandwidth) and therefore well-suited for benchmarking robustness and domain shift in interference classification, multi-task characterization, and downstream impact evaluation.

\section{EVALUATION}
\label{label_evaluation}

\subsection{Model Benchmark}

\begin{figure*}[!t]
    \centering
	\begin{minipage}[t]{0.325\linewidth}
        \centering
    	\includegraphics[trim=10 10 10 10, clip, width=1.0\linewidth]{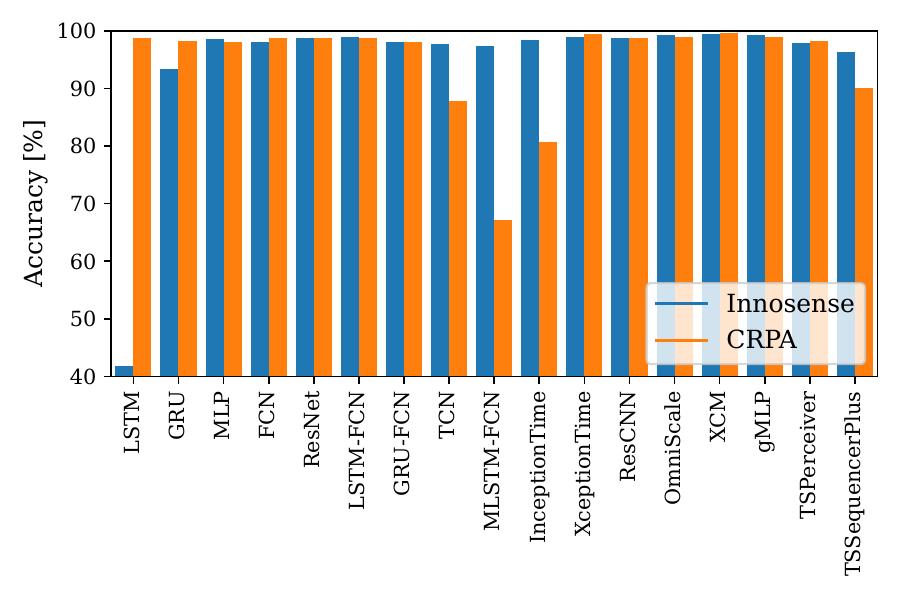}
        \vspace{-0.5cm}
        \subcaption{Interference classification accuracy [\%].}
        \label{figure_ML_benchmark1}
    \end{minipage}
    \hfill
	\begin{minipage}[t]{0.325\linewidth}
        \centering
    	\includegraphics[trim=10 10 10 10, clip, width=1.0\linewidth]{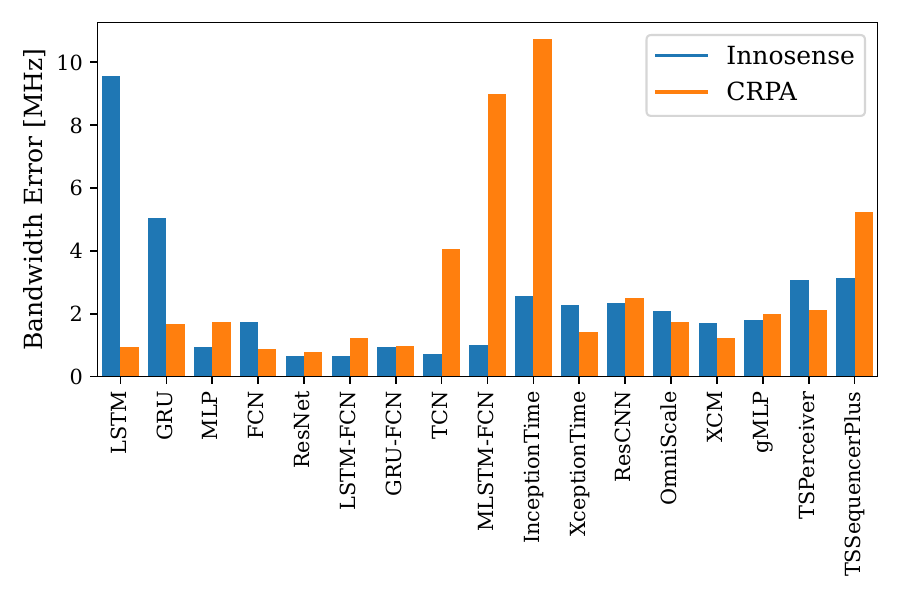}
        \vspace{-0.5cm}
        \subcaption{Bandwidth evaluation [MHz].}
        \label{figure_ML_benchmark2}
    \end{minipage}
    \hfill
	\begin{minipage}[t]{0.325\linewidth}
        \centering
    	\includegraphics[trim=10 10 10 10, clip, width=1.0\linewidth]{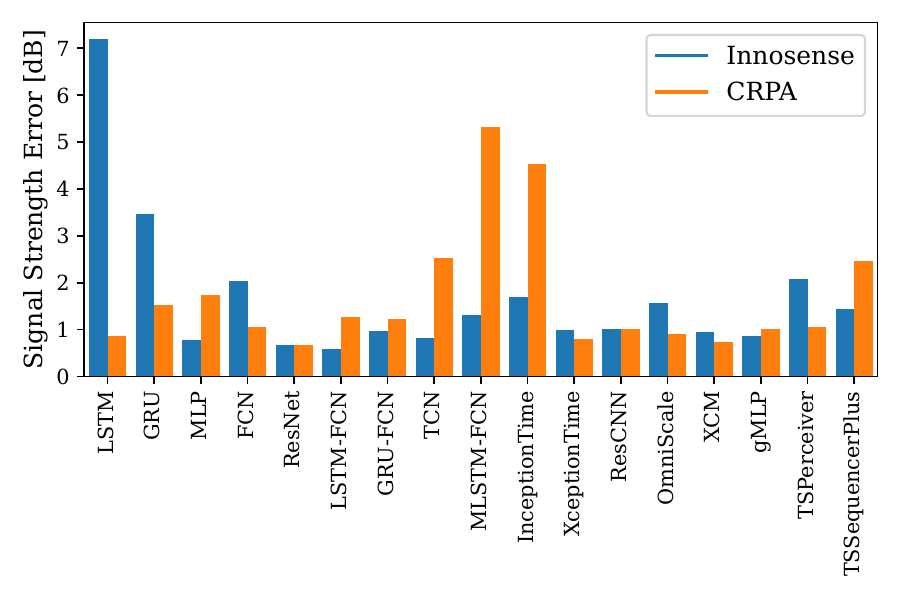}
        \vspace{-0.5cm}
        \subcaption{Signal strength evaluation [dB].}
        \label{figure_ML_benchmark3}
    \end{minipage}
    \vspace{-0.2cm}
    \caption{Benchmark of 17 time-series ML models for GNSS interference characterization using the Innosense and CRPA receiver datasets. The comparison reports (a) interference-type classification accuracy, (b) occupied-bandwidth estimation error, and (c) signal-strength estimation error. Higher values are better for classification accuracy, whereas lower values indicate better characterization performance for bandwidth and signal-strength estimation.}
    \label{figure_ML_benchmark}
    \vspace{-0.3cm}
\end{figure*}

Figure~\ref{figure_ML_benchmark} compares 17 time-series ML architectures based on tsai~\citep{tsai} for the joint interference-characterization task on the Innosense and CRPA receiver datasets. Overall, most convolutional, hybrid, and modern time-series architectures achieve high interference-type classification accuracy, with several models reaching close to or above $98\%$ for both receivers. In particular, FCN, ResNet, LSTM-FCN, GRU-FCN, XceptionTime, ResCNN, OmniScale, XCM, gMLP, and TSPerceiver show consistently strong classification performance, indicating that the time-series representations contain discriminative information for separating the considered interference families. However, the results show that classification accuracy alone is not sufficient to assess characterization quality. Some models with high type accuracy still exhibit large errors in bandwidth or signal-strength estimation, especially for the CRPA receiver. This is visible for InceptionTime and MLSTM-FCN, which show comparatively large CRPA bandwidth and signal-strength errors despite strong classification performance on at least one receiver. In contrast, ResNet provides one of the most balanced trade-offs across all three objectives, combining near-saturated classification accuracy with low bandwidth and signal-strength errors for both Innosense and CRPA. The recurrent baselines are less stable: LSTM performs poorly for Innosense classification and yields the largest Innosense bandwidth and signal-strength errors, while GRU improves classification but still shows elevated characterization errors. The CRPA results are generally more variable than the Innosense results, which is expected due to the shorter snapshot duration, array-specific preprocessing, and wider instantaneous bandwidth. Overall, the benchmark indicates that convolutional and hybrid time-series models are best suited for robust GNSS interference characterization, while the comparison across the three panels highlights the importance of evaluating type classification, occupied bandwidth, and signal strength jointly rather than selecting models based only on classification accuracy.

\subsection{Interference Characterization}

\begin{table*}[!t]
\caption{Summary of benchmark results for both receivers using an 80\,/\,20 train-test split. We evaluate a thresholding strategy in which very weak jammers below a predefined power level are excluded from training/evaluation. We compare two learning formulations: (1) a single-task classifier trained with a dedicated classification loss, and (2) a multi-task characterization model that jointly predicts interference type, occupied bandwidth, and signal strength using parallel loss terms. Confusion matrices for $[1]$--$[8]$ are presented in Figure~\ref{figure_confusion_matrix}.}
\label{tab_benchmark_results}
\vspace{-0.2cm}
\centering
\scriptsize
\setlength{\tabcolsep}{2.6pt}
\resizebox{\textwidth}{!}{%
\begin{tabular}{lcccccccccccc}
\toprule
\textbf{Receiver} & \multicolumn{3}{c}{\textbf{Training}} & \textbf{Test} & \multicolumn{4}{c}{\textbf{Results (Area 1)}} & \multicolumn{4}{c}{\textbf{Results (Area 2)}} \\
\cmidrule(lr){2-4}\cmidrule(lr){6-9}\cmidrule(lr){10-13}
& \textbf{Area} & \textbf{Band} & \textbf{Weak} & & \textbf{Type\textsuperscript{(1)}} & \textbf{Type\textsuperscript{(2)}} & \textbf{Bandwidth\textsuperscript{(2)}} & \textbf{Signal Strength\textsuperscript{(2)}} & \textbf{Type\textsuperscript{(1)}} & \textbf{Type\textsuperscript{(2)}} & \textbf{Bandwidth\textsuperscript{(2)}} & \textbf{Signal Strength\textsuperscript{(2)}} \\
& & & \textbf{Jammer} & & \textbf{[\%]} & \textbf{[\%]} & \textbf{[MHz]} & \textbf{[dB]} & \textbf{[\%]} & \textbf{[\%]} & \textbf{[MHz]} & \textbf{[dB]} \\
\midrule
Innosense & 1   & E1 & w/  & 80/20 & 99.56$^{[1]}$            & 99.49      & \,\,\,0.29 & \,\,\,1.18 & 37.19\,\,\,\,\,\,\, & 33.01\,\,\,\,\,\,\,      & 15.05      & 17.66 \\
Innosense & 1   & E1 & w/o & 80/20 & 99.58\,\,\,\,\,\,\,      & 99.64      & \,\,\,0.21 & \,\,\,0.48 & 21.86$^{[3]}$       & 22.24\,\,\,\,\,\,\,      & 13.54      & \,\,\,8.67 \\
Innosense & 1   & E5 & w/  & 80/20 & 99.35\,\,\,\,\,\,\,      & 99.34      & \,\,\,0.53 & \,\,\,1.12 & --                  & --                       & --         & -- \\
Innosense & 1   & E5 & w/o & 80/20 & 99.65\,\,\,\,\,\,\,      & 99.75      & \,\,\,0.30 & \,\,\,0.47 & --                  & --                       & --         & -- \\
Innosense & 2   & E1 & w/  & 80/20 & 14.65\,\,\,\,\,\,\,      & 14.91      & 14.95      & 16.81      & 97.08$^{[2]}$       & 90.63\,\,\,\,\,\,\,      & \,\,\,4.64 & \,\,\,3.73 \\
Innosense & 2   & E1 & w/o & 80/20 & 15.71\,\,\,\,\,\,\,      & 14.46      & 10.53      & 11.43      & 98.05\,\,\,\,\,\,\, & 97.70\,\,\,\,\,\,\,      & \,\,\,1.33 & \,\,\,0.70 \\
Innosense & 1+2 & E1 & w/  & 80/20 & 99.48\,\,\,\,\,\,\,      & 97.48      & \,\,\,0.52 & \,\,\,1.67 & 97.08\,\,\,\,\,\,\, & 88.99$^{[4]}$            & \,\,\,6.31 & \,\,\,2.65 \\
Innosense & 1+2 & E1 & w/o & 80/20 & 99.58\,\,\,\,\,\,\,      & 99.41      & \,\,\,0.35 & \,\,\,0.79 & 98.14\,\,\,\,\,\,\, & 98.06\,\,\,\,\,\,\,      & \,\,\,1.60 & \,\,\,0.79 \\
CRPA      & 1   & E1 & w/  & 80/20 & 99.53$^{[5]}$            & 98.26      & \,\,\,1.15 & \,\,\,4.07 & 17.01\,\,\,\,\,\,\, & 26.36\,\,\,\,\,\,\,      & 36.67      & 19.47 \\
CRPA      & 1   & E1 & w/o & 80/20 & 99.41\,\,\,\,\,\,\,      & 99.33      & \,\,\,0.37 & \,\,\,0.75 & \,\,\,4.59$^{[7]}$  & \,\,\,4.91\,\,\,\,\,\,\, & 32.00      & 29.25 \\
CRPA      & 2   & E1 & w/  & 80/20 & \,\,\,3.55\,\,\,\,\,\,\, & \,\,\,9.43 & 39.80      & 15.73      & 99.21\,\,\,\,\,\,\, & 90.20\,\,\,\,\,\,\,      & \,\,\,8.24 & \,\,\,6.64 \\
CRPA      & 2   & E1 & w/o & 80/20 & \,\,\,4.31\,\,\,\,\,\,\, & \,\,\,5.62 & 71.25      & 11.20      & 99.28\,\,\,\,\,\,\, & 81.41\,\,\,\,\,\,\,      & \,\,\,8.96 & \,\,\,3.23 \\
CRPA      & 1+2 & E1 & w/  & 80/20 & 99.50\,\,\,\,\,\,\,      & 98.10      & \,\,\,3.68 & \,\,\,2.75 & 99.21$^{[6]}$       & 91.89$^{[8]}$            & \,\,\,7.06 & \,\,\,6.42 \\
CRPA      & 1+2 & E1 & w/o & 80/20 & 99.38\,\,\,\,\,\,\,      & 96.16      & \,\,\,2.94 & \,\,\,2.09 & 99.01\,\,\,\,\,\,\, & 92.16\,\,\,\,\,\,\,      & \,\,\,5.03 & \,\,\,2.03 \\
\bottomrule
\end{tabular}%
}
\end{table*}

Table~\ref{tab_benchmark_results} summarizes benchmark performance for two receiver setups (Innosense and a $2 \times 2$-element CRPA) using an 80/20 train-test split, reporting results separately for Test Area 1 and Test Area 2 and for training on data from Area 1, Area 2, or the combined Areas 1+2, with and without an additional weak jammer (“w/”, “w/o”). For each configuration, it reports the modulation-type classification accuracy (Type [\%]) and (when available) errors for bandwidth [MHz] and signal strength estimation [dB]. Overall, training and testing within the same area yields very high type accuracies ($\approx 99\%$ in Area 1, and $\approx 97-99\%$ in Area 2), whereas cross-area generalization degrades substantially (e.g., models trained on Area 1 performing poorly on Area 2, and vice versa), highlighting a pronounced domain shift between the two test environments. Training on the combined Areas 1+2 improves robustness across both areas and enables consistent multi-task characterization, with bandwidth errors on the order of $\approx 0.35-0.52\,\text{MHz}$ for Innosense (E1) and larger ($\approx 2.9-3.7\,\text{MHz}$) for CRPA, while signal strength errors range from sub-dB to a few dB depending on the configuration.

\newcommand\w{0.245}
\begin{figure*}[!t]
    \centering
	\begin{minipage}[t]{\w\linewidth}
        \centering
    	\includegraphics[trim=16 12 18 10, clip, width=1.0\linewidth]{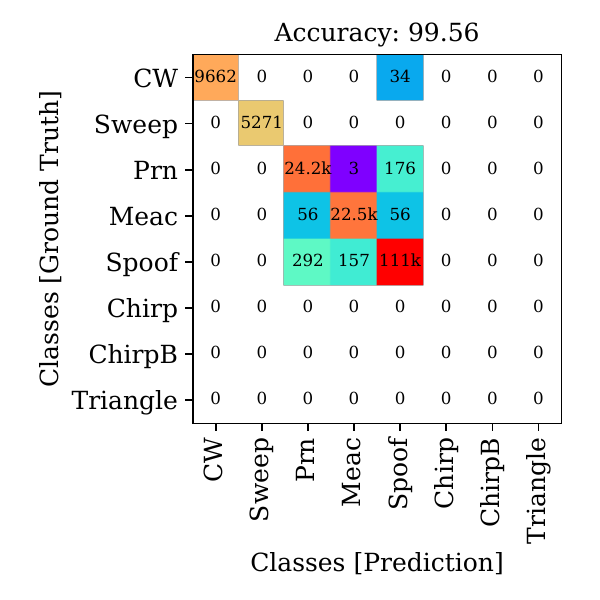}
        \subcaption{Innosense, Train Area 1, Test Area 1, E1, classification$^{[1]}$.}
        \label{figure_confusion_matrix1}
    \end{minipage}
    \hfill
	\begin{minipage}[t]{\w\linewidth}
        \centering
    	\includegraphics[trim=16 12 18 10, clip, width=1.0\linewidth]{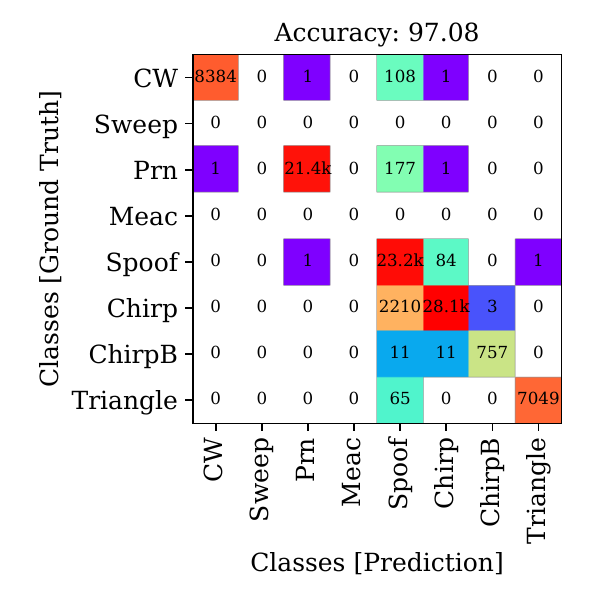}
        \subcaption{Innosense, Train Area 2, Test Area 2, E1, classification$^{[2]}$.}
        \label{figure_confusion_matrix2}
    \end{minipage}
    \hfill
	\begin{minipage}[t]{\w\linewidth}
        \centering
    	\includegraphics[trim=16 12 18 10, clip, width=1.0\linewidth]{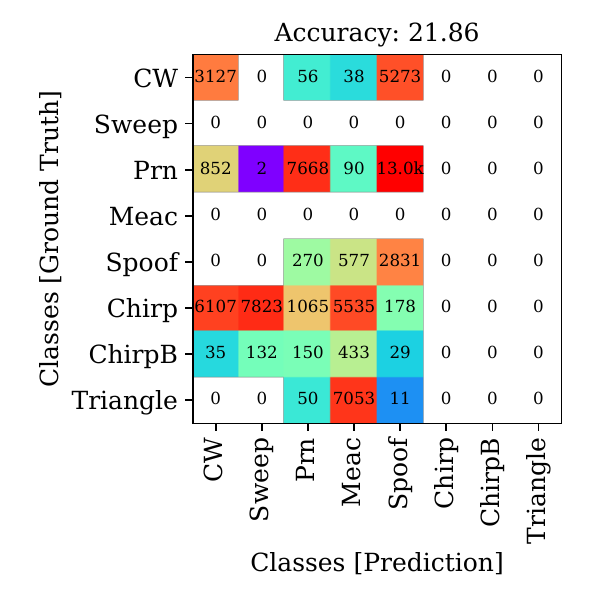}
        \subcaption{Innosense, Train Area 1, Test Area 2, E1, classification$^{[3]}$.}
        \label{figure_confusion_matrix3}
    \end{minipage}
    \hfill
	\begin{minipage}[t]{\w\linewidth}
        \centering
    	\includegraphics[trim=16 12 18 10, clip, width=1.0\linewidth]{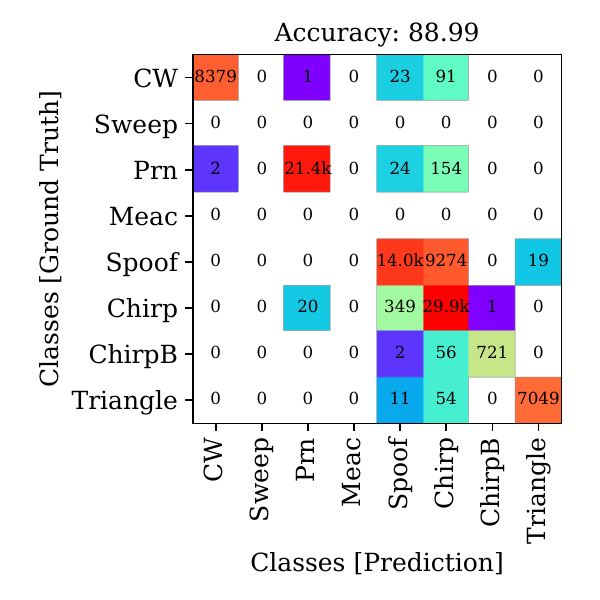}
        \subcaption{Innosense, Train Area 1+2, Test Area 2, E1, characterization$^{[4]}$.}
        \label{figure_confusion_matrix4}
    \end{minipage}
	\begin{minipage}[t]{\w\linewidth}
        \centering
    	\includegraphics[trim=16 12 18 10, clip, width=1.0\linewidth]{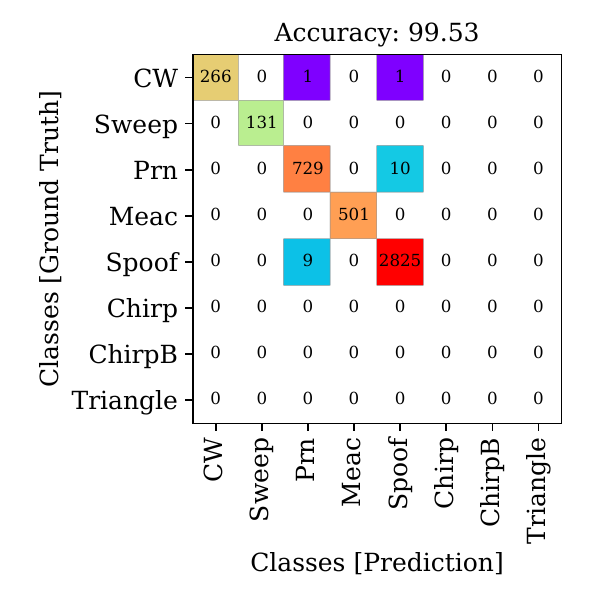}
        \subcaption{CRPA, Train Area 1, Test Area 1, E1, classification$^{[5]}$.}
        \label{figure_confusion_matrix5}
    \end{minipage}
    \hfill
	\begin{minipage}[t]{\w\linewidth}
        \centering
    	\includegraphics[trim=16 12 18 10, clip, width=1.0\linewidth]{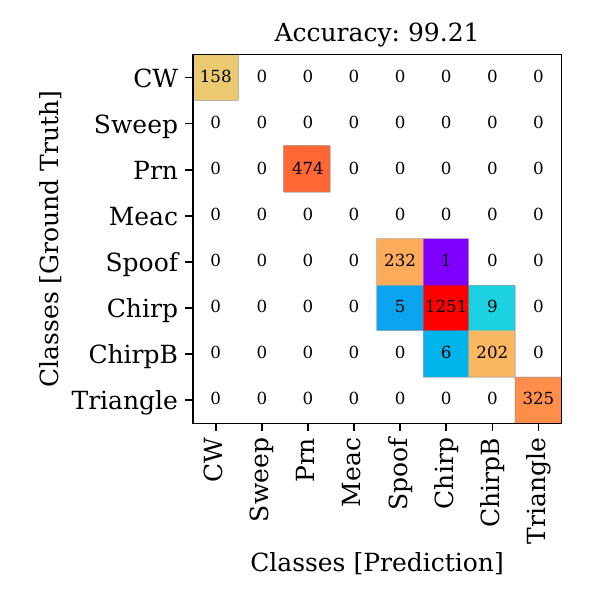}
        \subcaption{CRPA, Train Area 2, Test Area 2, E1, classification$^{[6]}$.}
        \label{figure_confusion_matrix6}
    \end{minipage}
    \hfill
	\begin{minipage}[t]{\w\linewidth}
        \centering
    	\includegraphics[trim=16 12 18 10, clip, width=1.0\linewidth]{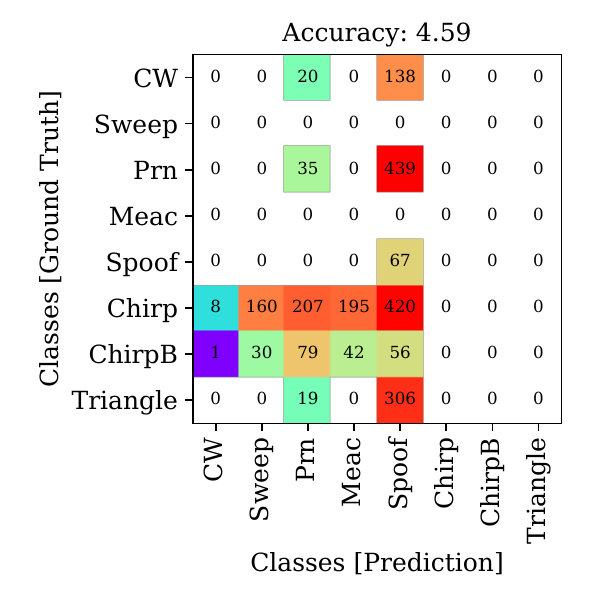}
        \subcaption{CRPA, Train Area 1, Test Area 2, E1, classification$^{[7]}$.}
        \label{figure_confusion_matrix7}
    \end{minipage}
    \hfill
	\begin{minipage}[t]{\w\linewidth}
        \centering
    	\includegraphics[trim=16 12 18 10, clip, width=1.0\linewidth]{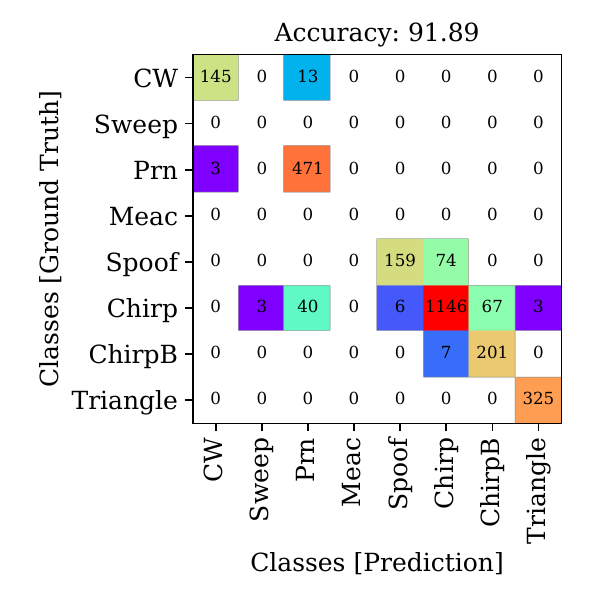}
        \subcaption{CRPA, Train Area 1+2, Test Area 2, E1, characterization$^{[8]}$.}
        \label{figure_confusion_matrix8}
    \end{minipage}
    \vspace{-0.2cm}
    \caption{Confusion matrices for different train/test scenarios. Results of $[1]$--$[8]$ correspond to entries in Table~\ref{tab_benchmark_results}.}
    \label{figure_confusion_matrix}
\end{figure*}

The confusion matrices in Figure~\ref{figure_confusion_matrix} provide additional insight into the benchmark results reported in Table~\ref{tab_benchmark_results}. For matched train/test conditions, the matrices are strongly diagonal for both receiver types, confirming that the models learn highly discriminative representations when training and evaluation are performed within the same area. This is visible for Innosense in Area~1 ($99.56\%$, Fig.~\ref{figure_confusion_matrix1}) and Area~2 ($97.08\%$, Fig.~\ref{figure_confusion_matrix2}), and similarly for the CRPA setup in Area~1 ($99.53\%$, Fig.~\ref{figure_confusion_matrix5}) and Area~2 ($99.21\%$, Fig.~\ref{figure_confusion_matrix6}). In these cases, most samples are assigned to the correct interference family, with only limited confusion between spectrally related classes. In contrast, the cross-area evaluations reveal a pronounced domain shift. When training on Area~1 and testing on Area~2, the classification accuracy drops substantially for both Innosense ($21.86\%$, Fig.~\ref{figure_confusion_matrix3}) and CRPA ($4.59\%$, Fig.~\ref{figure_confusion_matrix7}), producing strong off-diagonal structure in the confusion matrices. This indicates that models trained on one campaign area do not directly generalize to the different waveform catalogue, propagation environment, and receiver conditions of the other area. In particular, Area~2 contains several chirp-like and triangular interference types that are absent or underrepresented in Area~1, causing these samples to be mapped to the closest known classes such as PRN, spoofing, or other broadband patterns. Joint training on Area~1 and Area~2 substantially reduces this effect and restores a dominant diagonal structure. For example, Innosense achieves $88.99\%$ characterization accuracy on Area~2 (Fig.~\ref{figure_confusion_matrix4}), while CRPA reaches $91.89\%$ characterization accuracy under the same joint-training setting (Fig.~\ref{figure_confusion_matrix8}). The remaining errors are mainly concentrated between classes with similar time--frequency morphology, for example between chirp-related classes or between composite scenarios with overlapping spectral support. Overall, the confusion matrices show that the proposed models perform reliably under matched conditions, but that robust deployment requires exposure to the heterogeneous waveform and propagation conditions present across both test areas; combined-area training therefore provides a clear benefit for domain generalization and multi-task interference characterization.

\subsection{Cross-Dataset Generalization to Jammertest}

To quantify how well models trained on external (partly publicly available) datasets transfer to our field-recorded Jammertest measurements, we perform a cross-dataset evaluation in which training is conducted on one dataset at a time and testing is always performed on the fixed Jammertest test set (Table~\ref{tab:cross_dataset_results}). For the Innosense indoor datasets, proposed by \cite{heublein_feigl_crpa}, recorded at the Fraunhofer IIS L.I.N.K.~test and application center (same receiver family and comparable front-end chain), the model achieves classification accuracies ranging from $53.8\%$ to $83.2\%$ across amplifier configurations, indicating that receiver gain and indoor propagation (including multipath) materially affect the learned decision boundaries. The corresponding characterization accuracies are consistently lower (from $27.6\%$ to $50.4\%$), which is expected because the indoor training data contain a substantially finer label space (e.g., $\approx 310$ interference classes) and therefore require a more granular mapping to the coarser Jammertest taxonomy; this label-space mismatch and reduced per-class support after remapping particularly penalize the multi-attribute characterization task.

\begin{table}[!t]
\centering
\small
\setlength{\tabcolsep}{3.2pt}
\caption{Generalization to Jammertest: models trained on different datasets and evaluated on the Jammertest test set (Area 1+2, band E1).}
\label{tab:cross_dataset_results}
\vspace{-0.2cm}
\begin{tabular}{lcc}
\toprule
\textbf{Dataset (train)} & \textbf{Classification acc. [\%]} & \textbf{Characterization acc. [\%]} \\
\midrule
Innosense (indoor) amplifier 1   & 53.8 & 41.6 \\
Innosense (indoor) amplifier 2   & 83.2 & 50.4 \\
Innosense (indoor) amplifier 3   & 57.4 & 27.6 \\
Innosense (indoor) with repeater & 47.3 & 44.4 \\
Innosense (outdoor, open field)  & 79.1 & 82.2 \\
Sionna RT (simulation)           & 37.8 & -- \\
CRPA (indoor)                    & 45.0 & \,\,\,0.6 \\
CRPA (outdoor, open field)       & 67.0 & 22.5 \\
\bottomrule
\end{tabular}
\end{table}

Training on the Innosense open-field dataset~\citep{heublein_benschuh} yields the most robust transfer to Jammertest, with $79.1\%$ classification accuracy and $82.2\%$ characterization accuracy. Notably, characterization exceeds classification in this setting, which can occur when multiple fine-grained training classes map onto a single Jammertest class (class pooling): the effective number of target classes decreases while the number of samples per mapped class increases, improving the stability of the multi-task heads despite residual domain shift. In contrast, pretraining on Sionna ray-tracing data, proposed by \cite{wielenberg_heublein_sionna}, exhibits a pronounced sim-to-real gap: the resulting model achieves only $37.8\%$ classification accuracy on Jammertest, suggesting overfitting to simulation-specific channel, noise, and signal-generation statistics that do not match the real receiver and environment.

Finally, the CRPA-based transfer results are substantially weaker than the single-antenna Innosense transfer. Training on CRPA indoor data~\citep{heublein_feigl_crpa} reaches only $45.0\%$ classification accuracy and effectively fails on characterization ($0.6\%$), while CRPA open-field~\citep{heublein_benschuh} training improves classification to $67.0\%$ but still yields limited characterization accuracy ($22.5\%$). A key driver is the effective temporal resolution/observation constraints of the CRPA recordings (and associated preprocessing), which can suppress or smear discriminative temporal signatures; consequently, certain interference families (e.g., \textit{pulsed} components) may not be reliably observable in the available IQ snapshots, leading to systematic ambiguity for both type recognition and downstream attribute estimation. Overall, these results highlight that (i) domain shift across environments and receiver front ends is a dominant factor for generalization, (ii) label-space granularity and class-mapping strongly influence characterization accuracy, and (iii) simulation-only pretraining remains insufficient without additional domain adaptation or calibration to real receiver statistics.

\subsection{Evaluation of Interference Degradation-Effects}

\begin{figure*}[!t]
    \centering
	\begin{minipage}[t]{0.325\linewidth}
        \centering
    	\includegraphics[trim=42 10 42 10, clip, width=1.0\linewidth]{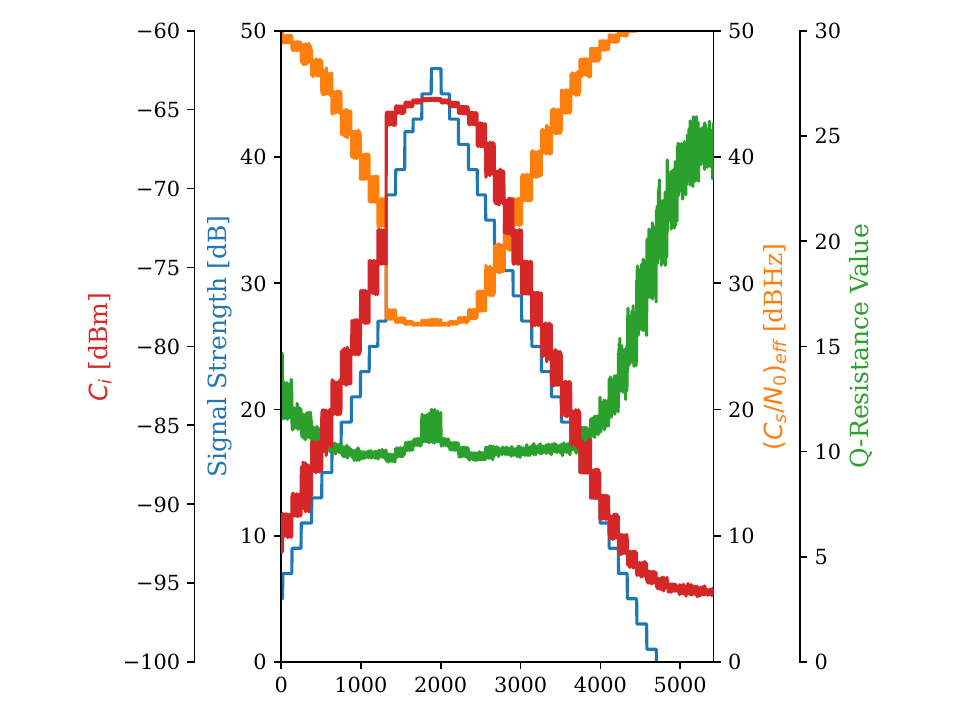}
        \subcaption{F8.1, PRN, 20MHz.}
        \label{image_ssc_evaluation1}
    \end{minipage}
    \hfill
	\begin{minipage}[t]{0.325\linewidth}
        \centering
    	\includegraphics[trim=40 10 42 10, clip, width=1.0\linewidth]{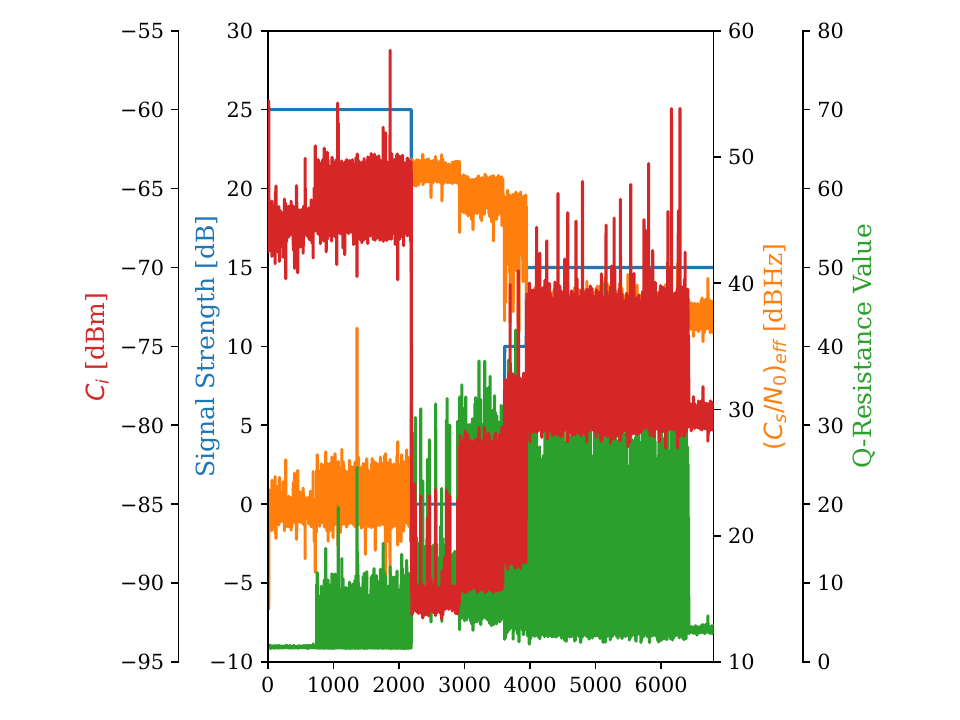}
        \subcaption{S, Spoofing.}
        \label{image_ssc_evaluation2}
    \end{minipage}
    \hfill
	\begin{minipage}[t]{0.325\linewidth}
        \centering
    	\includegraphics[trim=42 10 42 10, clip, width=1.0\linewidth]{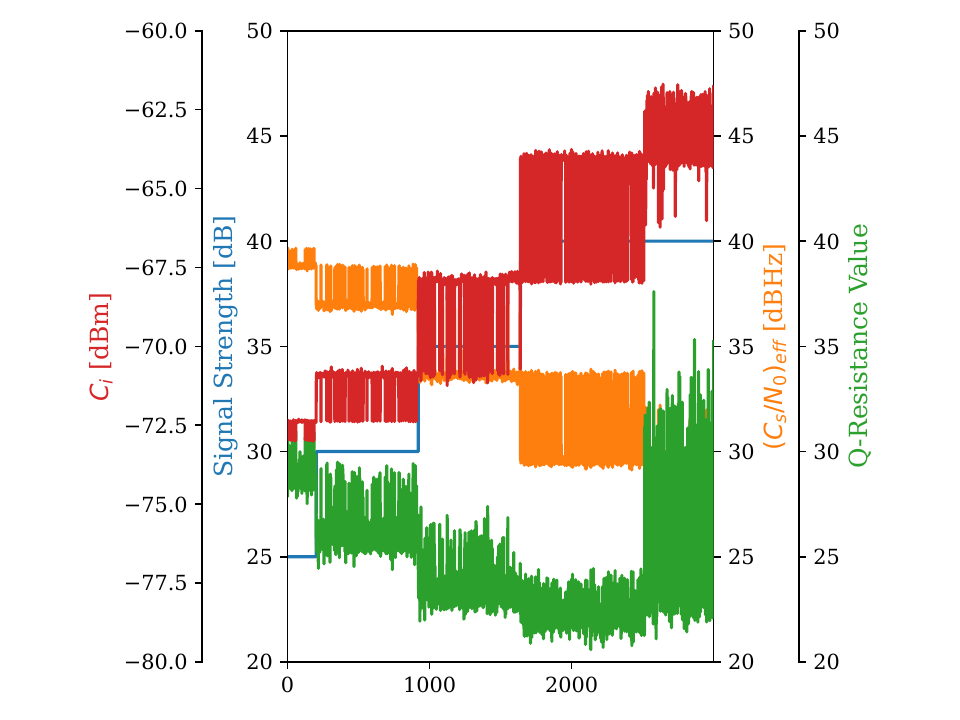}
        \subcaption{F1.1, Meaconing.}
        \label{image_ssc_evaluation3}
    \end{minipage}
    \vspace{-0.2cm}
    \caption{Example of receiver-aware SSC impact evaluation under stepped-power interference scenarios. The red trace shows the estimated interference power $C_i$ (dBm), the blue trace the predicted interference signal strength (dB), the orange trace the resulting effective carrier-to-noise density ratio $(C_s/N_0)_{\mathrm{eff}}$ (dB-Hz), and the green trace the derived jamming-resistance value $Q$.}
    \label{image_ssc_evaluation}
\end{figure*}

Figure~\ref{image_ssc_evaluation} illustrates how the proposed receiver-aware SSC pipeline disentangles \emph{power} and \emph{spectral overlap} effects to obtain navigation-relevant degradation. Across all three scenarios, the estimated interference power $C_i$ (red) increases in steps or phases and is mirrored by the predicted signal-strength trace (blue), while the effective carrier-to-noise density ratio $(C_s/N_0)_{\mathrm{eff}}$ (orange) decreases accordingly, consistent with the mapping in Eq.~\ref{eq:cn0eff_est}. The green curve shows the derived jamming-resistance factor $Q$, which is inversely related to the SSC via $Q \approx (R_c\,\kappa_{is})^{-1}$ (Eq.~\ref{eq:Q_est}) and therefore summarizes how \textit{spectrally similar} the interference is to the desired GNSS signal after receiver filtering (Eq.~\ref{eq:ssc-receiver-aware-method}). In the PRN case (Fig.~\ref{image_ssc_evaluation1}), a pronounced rise in $C_i$ coincides with a strong reduction in $(C_s/N_0)_{\mathrm{eff}}$, while $Q$ remains comparatively low and only recovers when the interference becomes less effective (smaller overlap / reduced in-band energy), indicating that the degradation is driven by both increased power and persistently large spectral overlap. In the spoofing example (Fig.~\ref{image_ssc_evaluation2}), $(C_s/N_0)_{\mathrm{eff}}$ exhibits abrupt changes despite similar power levels, and the corresponding variations in $Q$ reveal that changes in the spoofing waveform (and thus $\kappa_{is}$) can substantially alter impact even when $C_i$ is not the only driver. In the meaconing scenario (Fig.~\ref{image_ssc_evaluation3}), the stepwise increase of $C_i$ yields a near-monotonic decrease of $(C_s/N_0)_{\mathrm{eff}}$, while $Q$ stays relatively stable over most of the interval, indicating an approximately constant overlap regime where impact is predominantly governed by power. Overall, these examples highlight a key advantage of our method: by estimating $C_i$ directly from IQ samples (Eq.~\ref{eq:Ci_est}) and computing a receiver-aware overlap metric $\kappa_{is}$ (Eqs.~\ref{eq:diff_psd}--\ref{eq:ssc_discrete}), the pipeline provides an interpretable, receiver-specific prediction of $(C_s/N_0)_{\mathrm{eff}}$ degradation that remains comparable across fundamentally different interference types without requiring jammer-specific models.

\subsection{Multi-Source Interference Detection}

\paragraph{Bounding Boxes.} Figure~\ref{figure_yolo_rfdetr_results} compares YOLOv8s and RF-DETR bounding-box predictions on representative GNSS spectrogram snapshots for multi-source interference detection. The upper row (Figures~\ref{figure_yolo_rfdetr_results}a--f) shows YOLOv8s detections, where the blue boxes localize interference components in the time--frequency plane and the associated confidence scores indicate the reliability of each prediction. YOLOv8s provides fine-grained detections and is able to separate multiple simultaneous components, including thin narrowband structures, wider interference regions, and partially overlapping emissions. This is particularly visible in examples with several stacked or intersecting spectral components, where the detector assigns separate boxes to individual interference traces and, in some cases, explicitly labels overlap regions. The lower row (Figures~\ref{figure_yolo_rfdetr_results}g--l) shows the corresponding RF-DETR predictions. Compared with YOLOv8s, RF-DETR produces fewer and more conservative detections: it reliably captures dominant wideband or high-energy interference regions, but is less sensitive to very thin or closely spaced spectral components. This behavior suggests a trade-off between fine-grained localization and robust region-level detection. YOLOv8s is advantageous when individual interferers must be separated precisely for per-source characterization, while RF-DETR provides compact detections of dominant interference activity and may reduce over-segmentation in spectrograms with broad emissions. Overall, the results demonstrate that object-detection models can transfer the multi-source interference problem into a structured time--frequency localization task, enabling each detected component to be processed independently for subsequent modulation-type classification, occupied-bandwidth estimation, and interference-power assessment.

\newcommand\z{0.162}
\begin{figure*}[!t]
\captionsetup[subfigure]{font=scriptsize}
    \centering
	\begin{minipage}[t]{\z\linewidth}
        \centering
    	\includegraphics[trim=11 11 11 11, clip, width=1.0\linewidth]{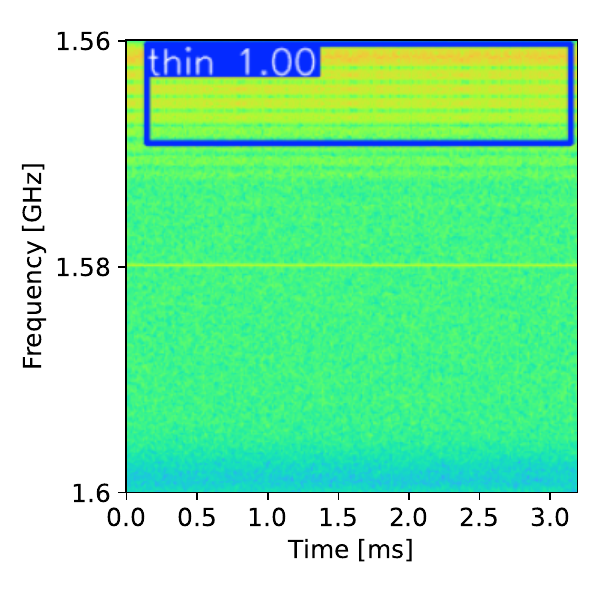}
        \subcaption{YOLOv8s, example 1.}
        \label{figure_yolo_rfdetr_results1}
    \end{minipage}
    \hfill
	\begin{minipage}[t]{\z\linewidth}
        \centering
    	\includegraphics[trim=11 11 11 11, clip, width=1.0\linewidth]{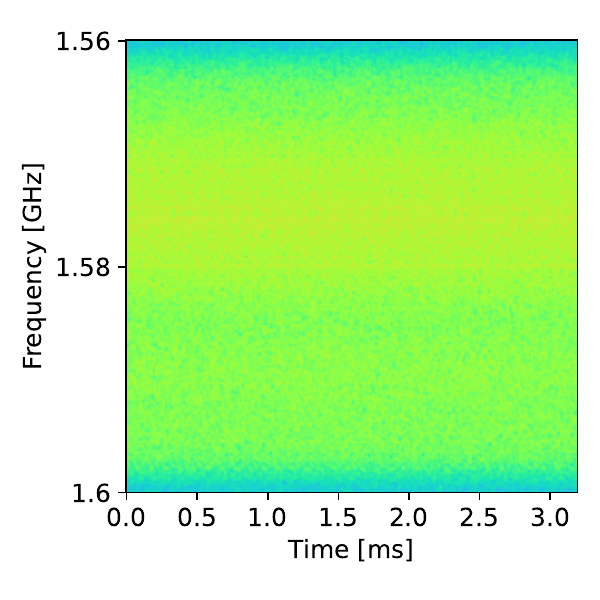}
        \subcaption{YOLOv8s, example 2.}
        \label{figure_yolo_rfdetr_results2}
    \end{minipage}
    \hfill
	\begin{minipage}[t]{\z\linewidth}
        \centering
    	\includegraphics[trim=11 11 11 11, clip, width=1.0\linewidth]{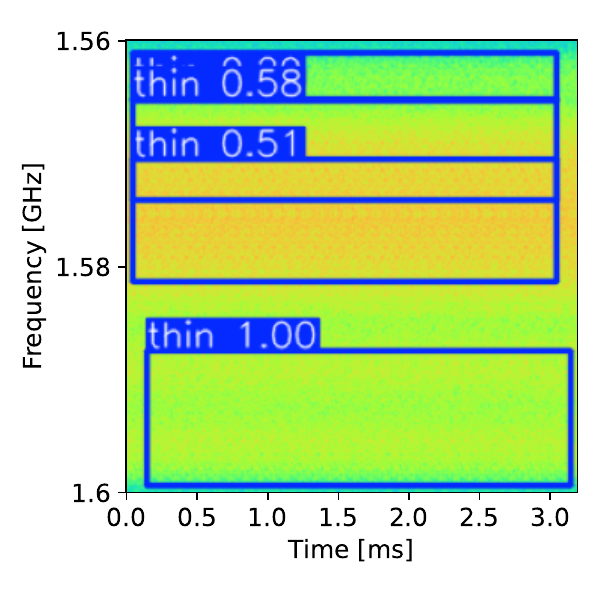}
        \subcaption{YOLOv8s, example 3.}
        \label{figure_yolo_rfdetr_results3}
    \end{minipage}
    \hfill
	\begin{minipage}[t]{\z\linewidth}
        \centering
    	\includegraphics[trim=11 11 11 11, clip, width=1.0\linewidth]{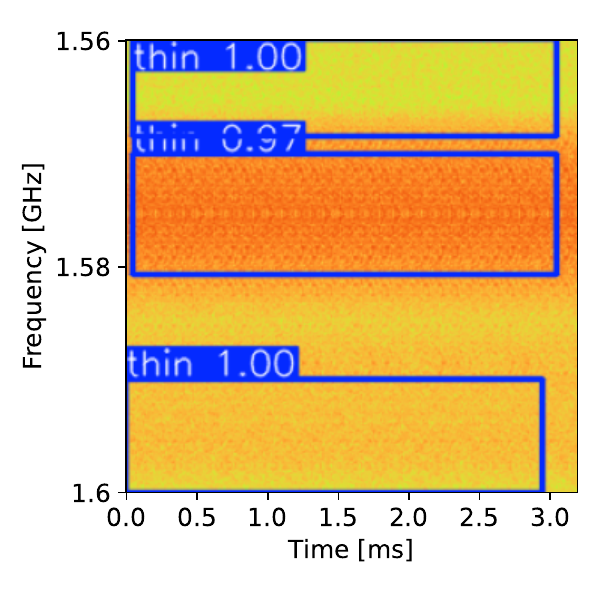}
        \subcaption{YOLOv8s, example 4.}
        \label{figure_yolo_rfdetr_results4}
    \end{minipage}
    \hfill
	\begin{minipage}[t]{\z\linewidth}
        \centering
    	\includegraphics[trim=11 11 11 11, clip, width=1.0\linewidth]{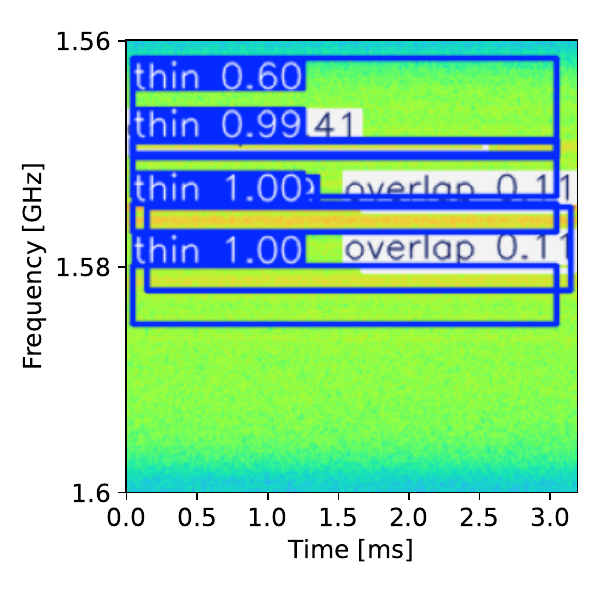}
        \subcaption{YOLOv8s, example 5.}
        \label{figure_yolo_rfdetr_results5}
    \end{minipage}
    \hfill
	\begin{minipage}[t]{\z\linewidth}
        \centering
    	\includegraphics[trim=11 11 11 11, clip, width=1.0\linewidth]{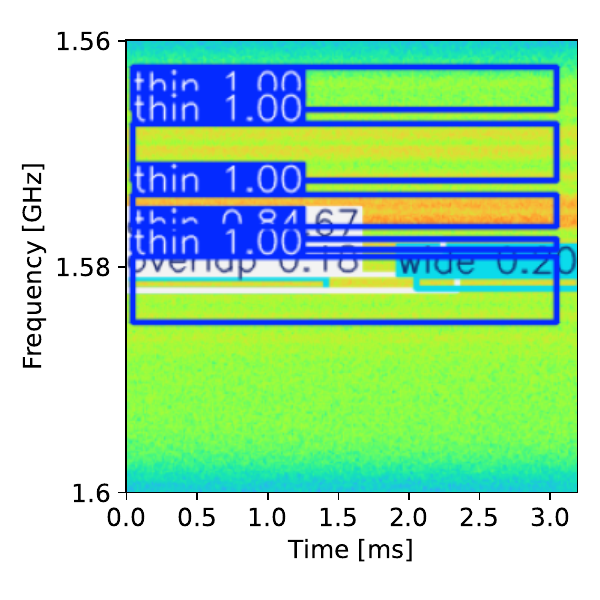}
        \subcaption{YOLOv8s, example 6.}
        \label{figure_yolo_rfdetr_results6}
    \end{minipage}
	\begin{minipage}[t]{\z\linewidth}
        \centering
    	\includegraphics[trim=11 11 11 11, clip, width=1.0\linewidth]{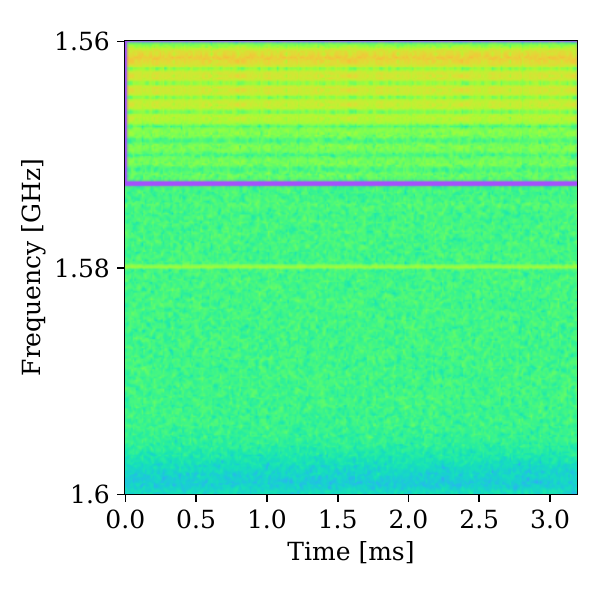}
        \subcaption{RF-DETR, example 1.}
        \label{figure_yolo_rfdetr_results7}
    \end{minipage}
    \hfill
	\begin{minipage}[t]{\z\linewidth}
        \centering
    	\includegraphics[trim=11 11 11 11, clip, width=1.0\linewidth]{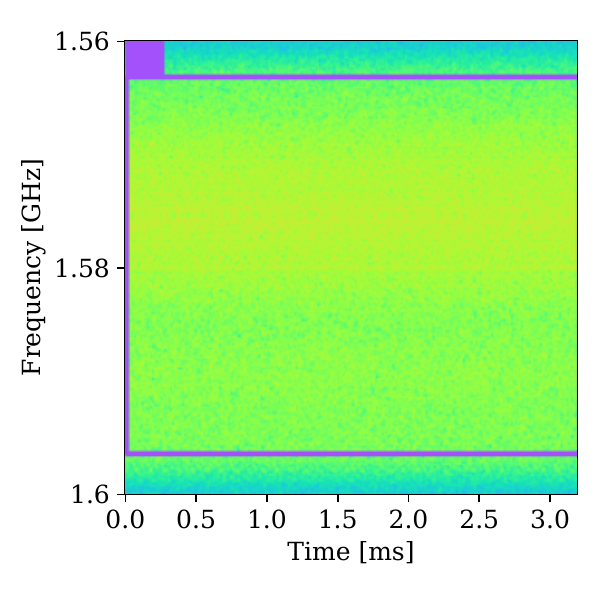}
        \subcaption{RF-DETR, example 2.}
        \label{figure_yolo_rfdetr_results8}
    \end{minipage}
    \hfill
	\begin{minipage}[t]{\z\linewidth}
        \centering
    	\includegraphics[trim=11 11 11 11, clip, width=1.0\linewidth]{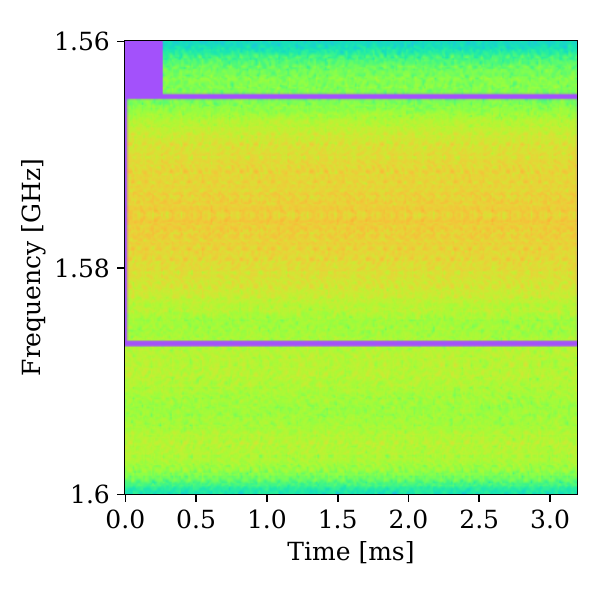}
        \subcaption{RF-DETR, example 3.}
        \label{figure_yolo_rfdetr_results9}
    \end{minipage}
    \hfill
	\begin{minipage}[t]{\z\linewidth}
        \centering
    	\includegraphics[trim=11 11 11 11, clip, width=1.0\linewidth]{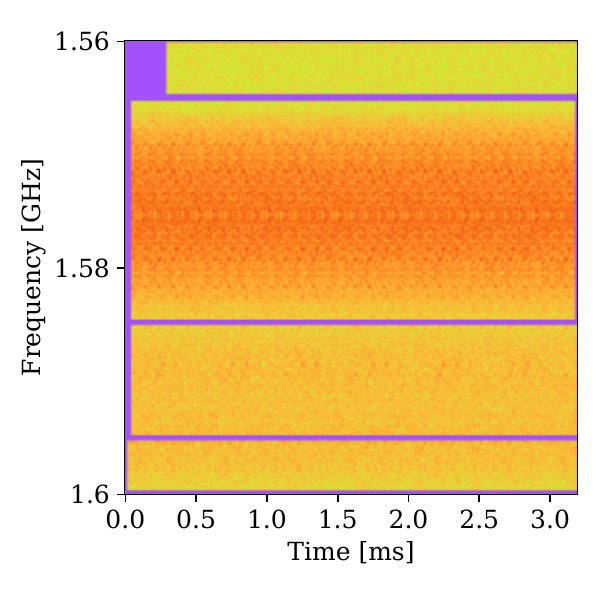}
        \subcaption{RF-DETR, example 4.}
        \label{figure_yolo_rfdetr_results10}
    \end{minipage}
    \hfill
	\begin{minipage}[t]{\z\linewidth}
        \centering
    	\includegraphics[trim=11 11 11 11, clip, width=1.0\linewidth]{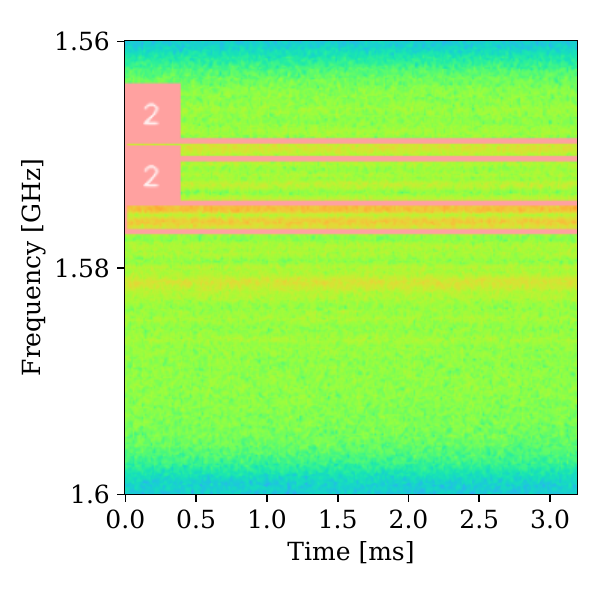}
        \subcaption{RF-DETR, example 5.}
        \label{figure_yolo_rfdetr_results11}
    \end{minipage}
    \hfill
	\begin{minipage}[t]{\z\linewidth}
        \centering
    	\includegraphics[trim=11 11 11 11, clip, width=1.0\linewidth]{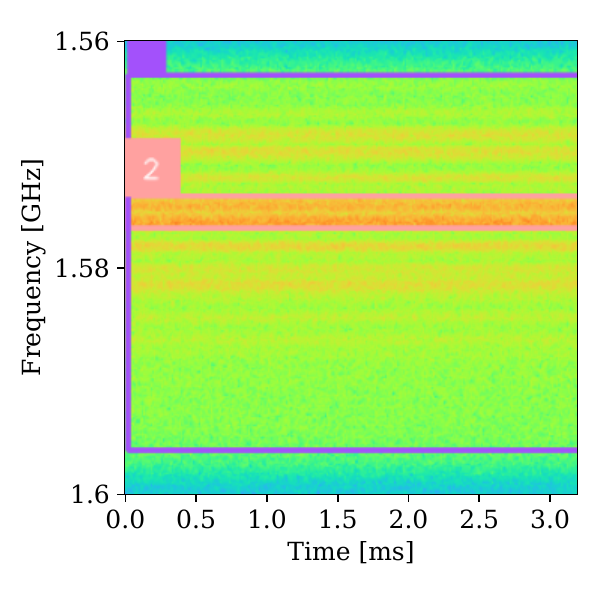}
        \subcaption{RF-DETR, example 6.}
        \label{figure_yolo_rfdetr_results12}
    \end{minipage}
    \vspace{-0.2cm}
    \caption{YOLOv8s (a to f) and RF-DETR (g to l) bounding-box predictions on GNSS spectrograms for multi-source interference detection. Each detected box delineates an individual interference component within the time-frequency plane, enabling subsequent per-source characterization (e.g., modulation type, occupied bandwidth, and interference power).}
    \label{figure_yolo_rfdetr_results}
\end{figure*}

\begin{figure*}[!t]
    \centering
	\begin{minipage}[t]{0.34\linewidth}
        \centering
    	\includegraphics[trim=10 10 10 10, clip, width=1.0\linewidth]{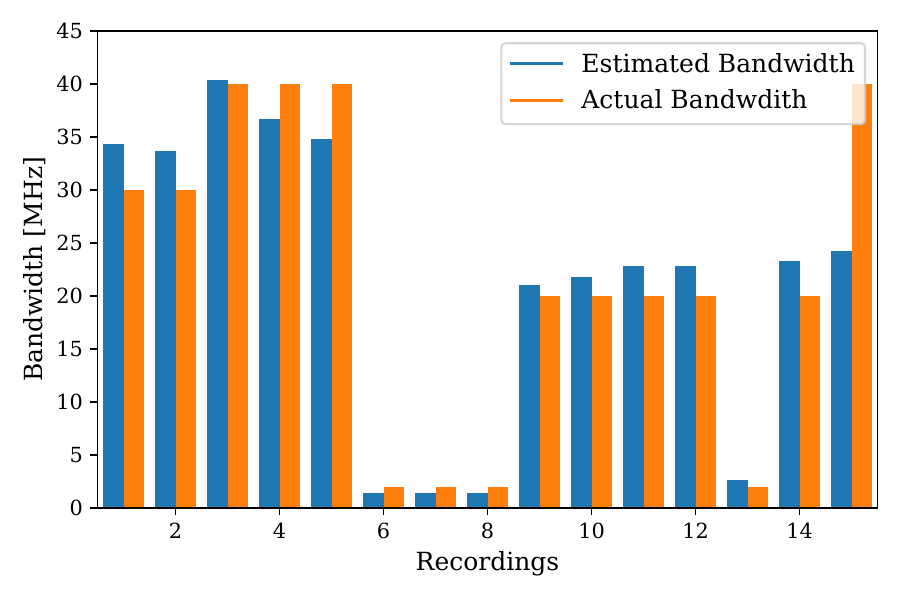}
        \vspace{-0.6cm}
        \caption{Bandwidth predictions from RF-DETR.}
        \label{figure_rf-detr_bandwidth}
    \end{minipage}
    \hfill
	\begin{minipage}[t]{0.15\linewidth}
        \centering
    	\includegraphics[trim=30 0 30 0, clip, width=1.0\linewidth]{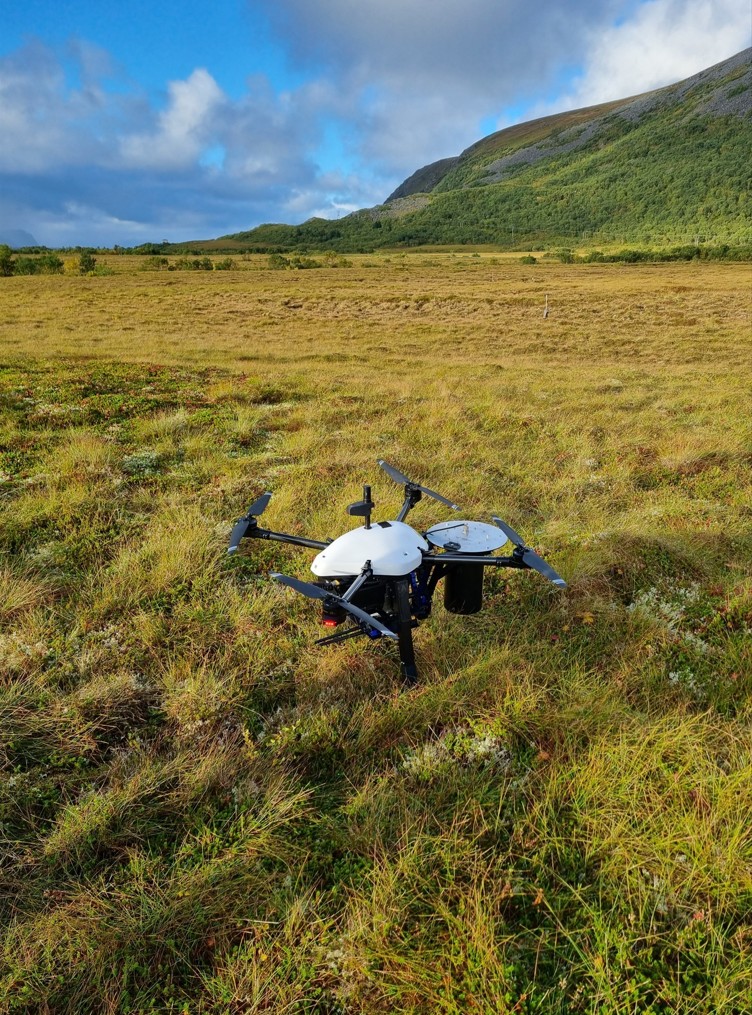}
        \vspace{-0.6cm}
        \caption{Drone setup.}
        \label{figure_direction_finding1}
    \end{minipage}
    \hfill
	\begin{minipage}[t]{0.23\linewidth}
        \centering
    	\includegraphics[trim=9 7 9 7, clip, width=1.0\linewidth]{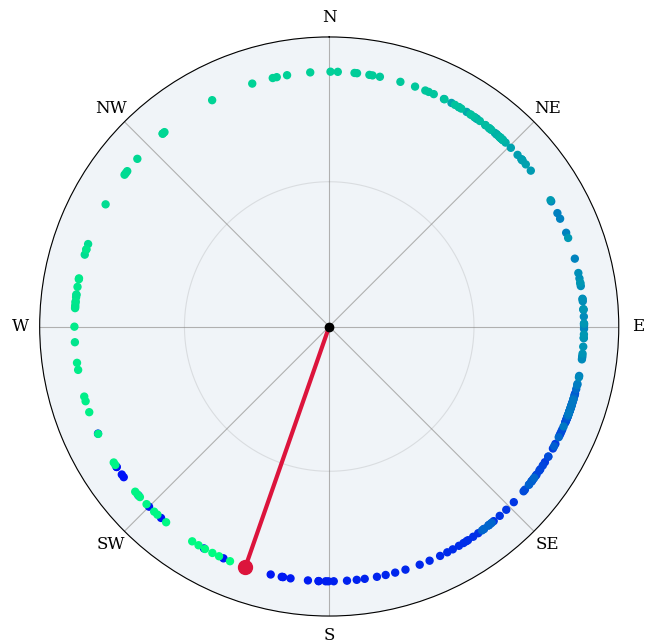}
        \vspace{-0.6cm}
        \caption{Direction finding of a mobile jammer on a drone.}
        \label{figure_direction_finding2}
    \end{minipage}
    \hfill
	\begin{minipage}[t]{0.23\linewidth}
        \centering
    	\includegraphics[trim=10 10 10 10, clip, width=1.0\linewidth]{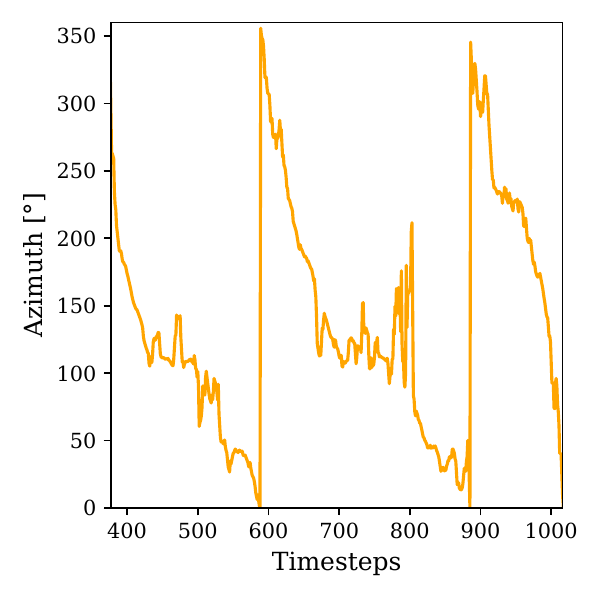}
        \vspace{-0.6cm}
        \caption{Azimuth prediction for direction finding.}
        \label{figure_direction_finding3}
    \end{minipage}
\end{figure*}

\paragraph{Bandwidth Estimation from Bounding Boxes.} Figure~\ref{figure_rf-detr_bandwidth} evaluates whether the RF-DETR bounding boxes provide a physically meaningful estimate of occupied interference bandwidth. Overall, the estimated bandwidth follows the nominal bandwidth labels well across the different recordings, indicating that the vertical extent of the predicted boxes captures the dominant spectral support of the interference. Narrowband events with very small bandwidths are estimated close to the ground-truth values, demonstrating that RF-DETR does not simply predict broad generic interference regions but can localize compact spectral components. For the $20\,\text{MHz}$-class recordings, the predicted bandwidths are slightly larger than the nominal labels, suggesting a mild overestimation of the spectral support, which can occur when the detector includes transition regions, spectral leakage, or low-power side components around the main interference band. For wider $30$--$40\,\text{MHz}$ events, the estimates are generally in the correct range, but some deviations are visible: several $40\,\text{MHz}$ recordings are underestimated, and the final recording shows a larger discrepancy, where the predicted box covers only part of the nominal bandwidth. This indicates that very wide or weakly bounded interference emissions remain more challenging, especially when the visible spectrogram energy does not occupy the full nominal transmitter bandwidth or when the detector focuses on the strongest subregion. Overall, the results confirm that RF-DETR bounding boxes are suitable not only for detecting interference presence but also for deriving a first-order bandwidth estimate that can support downstream per-source characterization. However, the remaining under- and over-estimation errors show that bounding-box-based bandwidth estimation should be interpreted as an estimate of the visible receiver-observed spectral occupancy rather than a perfect reconstruction of the nominal transmitter bandwidth.

\subsection{MUSIC-based Direction Finding} 

We evaluate direction finding with the CRPA by placing the $2 \times 2$ patch array at the center of Area~1. The jammer was mounted on a drone (see Figure~\ref{figure_direction_finding1}) that moved around the array while targeting three predefined fixed points. Since no precise ground-truth drone trajectory is available, the MUSIC~\citep{schmidt}-based azimuth estimates are analyzed qualitatively. Figure~\ref{figure_direction_finding2} shows the predicted direction sequence on a polar plot, where the color encodes temporal progression. The estimates form a coherent arc around the CRPA rather than random directions, indicating that the array captures the dominant angular motion of the mobile jammer. The smooth temporal color transition further suggests that consecutive IQ snapshots yield consistent angle estimates and that MUSIC can track the jammer movement over time, although the absolute trajectory cannot be quantitatively validated without reference labels. Figure~\ref{figure_direction_finding3} shows the azimuth predictions over time for several drone flights. The resulting curves exhibit structured, flight-dependent angular patterns with repeated rising and falling segments, which is consistent with the drone moving between different viewing directions around the antenna. Abrupt changes and local fluctuations likely originate from multipath, drone attitude changes, varying jammer-to-array geometry, and the limited aperture of the $2 \times 2$ CRPA. Overall, the results indicate that the recorded CRPA IQ data contain exploitable spatial information for mobile-jammer direction finding and motivate future experiments with synchronized drone-position ground truth for quantitative AoA evaluation.

\section{CONCLUSION}
\label{label_conclusion}

\paragraph{Summary.} This work presented a real-world GNSS interference monitoring dataset and an end-to-end evaluation pipeline based on measurements collected during \texttt{Jammertest\,2025}. The dataset combines synchronized recordings from two outdoor test areas and two complementary receiver setups, namely a compact Innosense receiver and a $2\times2$ CRPA array, and covers a broad range of jamming, spoofing, meaconing, chirp, PRN, CW, sweep, and multi-emitter scenarios. On this dataset, we evaluated time-series ML models for interference characterization, including waveform-type classification, occupied-bandwidth estimation, and signal-strength prediction. The results show that high performance is achievable under matched train/test conditions, with classification accuracies close to $99\%$ for both Innosense and CRPA. At the same time, the strong performance drops under cross-area testing demonstrate that real-world GNSS interference data exhibit pronounced domain shifts caused by changes in environment, waveform catalogue, receiver configuration, and propagation conditions. Joint training on Area~1 and Area~2 substantially mitigates this effect and enables robust multi-task characterization, particularly when weak jammer events are excluded. Beyond classification, the SSC-based impact evaluation links estimated interference power and receiver-aware spectral overlap to navigation-relevant degradation in $(C_s/N_0)_{\mathrm{eff}}$, while the multi-source detection experiments with YOLOv8s and RF-DETR show that spectrogram object detectors can localize individual interference components and provide physically meaningful bandwidth estimates from bounding boxes. Finally, the MUSIC-based drone experiment indicates that the CRPA recordings also contain exploitable spatial information for mobile-jammer direction finding.

\paragraph{Significance.} The significance of this work lies in connecting real-world GNSS interference measurements with an impact-aware monitoring pipeline. Rather than addressing detection or classification in isolation, the proposed approach jointly considers interference type, occupied bandwidth, signal strength, multi-source localization, and receiver-aware degradation through SSC-based analysis. The results show that high within-area accuracy does not necessarily imply robustness under realistic deployment conditions, as strong cross-area domain shifts remain visible. Therefore, the released Jammertest dataset and the proposed evaluation pipeline provide a practical benchmark for developing interference monitoring methods that are both data-driven and physically interpretable.


\section*{Acknowledgements}

This work has been carried out within the DARCII project, funding code 50NA2401, the PaiL project, funding code 50NP2506, and the SAIP project, funding code 50NP2603A, sponsored by the German Federal Ministry for Economic Affairs and Climate Action (BMWK) and the German Federal Ministry for Transport (BMV), and supported by the German Space Agency at DLR, the Bundesnetzagentur (BNetzA), and the Federal Agency for Cartography and Geodesy (BKG).

\bibliographystyle{apalike}
\bibliography{ION_GNSS}

\end{document}